\documentclass[fleqn,usenatbib]{mnras}

\usepackage{newtxtext,newtxmath}

\usepackage[T1]{fontenc}
\usepackage{ae,aecompl}

\usepackage{bm}
\let\vec\bm%

\usepackage{mleftright}
\let\left\mleft%
\let\right\mright%

\usepackage{graphicx}
\usepackage{amsmath}

\usepackage[normalem]{ulem}

\usepackage{notation}

\usepackage[dvipsnames]{xcolor}

\title[Analytic cosmological MHD code tests]{Hydromagnetic waves in an expanding universe -- cosmological MHD code tests using analytic solutions
}

\author[Thomas Berlok]{
Thomas Berlok$^{1}$\thanks{E-mail: tberlok@aip.de}
\\
$^{1}$Leibniz-Institut f{\"u}r Astrophysik Potsdam (AIP), An der Sternwarte 16, D-14482
 Potsdam, Germany
}

\date{Accepted XXX. Received YYY; in original form ZZZ}

\pubyear{2021}

\usepackage[dvipsnames]{xcolor}

\begin{document}
\label{firstpage}
\pagerange{\pageref{firstpage}--\pageref{lastpage}}
\maketitle

\begin{abstract}
We describe how analytic solutions for linear hydromagnetic waves can be used
for testing cosmological magnetohydrodynamic (MHD) codes. We start from the
comoving MHD equations and derive analytic solutions for the amplitude evolution
of linear hydromagnetic waves in a matter-dominated, flat Einstein-de-Sitter
(EdS) universe. The waves considered are comoving, linearly polarized Alfvén
waves and comoving, magnetosonic (fast) waves modified by self-gravity. The
solution for compressible waves is found for a general adiabatic index and we
consider the limits of hydrodynamics without self-gravity in addition to the
full solution. In addition to these analytic solutions, the linearized equations
are solved numerically for a $\Lambda$CDM cosmology. We use the analytic and
numeric solutions to compare with results obtained using the cosmological MHD
code \textsc{arepo} and find good agreement when
using a sufficient number of grid points. We interpret the numerical damping
clearly evident in simulations with few grid points by further deriving the
Alfvén wave solution including physical Navier-Stokes viscosity.
A comparison between Alfvén wave simulations and theory reveals that the
dissipation can be described by a numerical viscosity coefficient
$\eta_\mathrm{num} \propto a^{-5/2}$ where $a$ is the scale factor.  We envision that our
examples could be useful when developing a new cosmological MHD code or for
regression testing of existing codes.
\end{abstract}



\section{Introduction} \label{sec:intro}

Cosmological computer codes play an increasingly central role in theoretical
astrophysics. In their most basic state, these codes typically couple a
gravitational solver for dark matter with a hydrodynamic description of the
baryonic gas (e.g.
\citealt{Evrard1988,Springel2005,Teyssier2002,Springel2010,Bryan2014}). In
addition, many modern cosmological computer simulations also include physical
processes such as cooling, star formation and feedback from supernovae, stellar
winds and AGN (see \citealt{Vogelsberger2020} for a review and an extensive
list of physical processes typically included). While some of these physical
processes are included with phenomenological implementations, several different
codes have nevertheless managed to produce galaxies with realistic properties
\citep{Somerville2015}.

In recent years, many cosmological simulations have also begun to include
magnetic fields by solving the equations of ideal magnetohydrodynamics (MHD, see e.g.
\citealt{Freidberg2014}).
Cosmological codes with this capability include
\textsc{CosmoMHD} \citep{Li2008}, \textsc{MHD GADGET} \citep{Dolag2009},
\textsc{enzo} \citep{Collins2010,Bryan2014}, \textsc{arepo} \citep{Springel2010,Pakmor2013}, \textsc{gizmo} \citep{Hopkins2016}, \textsc{ramses} \citep{Dubois2008,Rieder2017}, \textsc{wombat} \citep{Mendygral2017},
\textsc{GCMHD++} \citep{Barnes2018}, and
\textsc{masclet} \citep{Quilis2020}.
These cosmological codes use comoving coordinates which follow the expansion of the universe. The transformation to comoving
coordinates transforms the standard MHD equations into the comoving MHD equations. The comoving version differs from the standard version by having factors of $a$
and $\dot{a}$ where $a$ is the scale factor and $\dot{a}$ is its time derivative. Despite this difference, only minor modifications of the algorithms used for standard MHD are required to solve the comoving MHD equations (e.g. \citealt{Pakmor2013}).

Examples of simulations performed with comoving MHD codes include
'universe-in-a-box' simulations (e.g.
\citealt{Marinacci2018,Hopkins2018,Katz2021,Garaldi2021}) and
'zoom' simulations which for instance aim to understand the role of magnetic fields in isolated or merging galaxies (e.g.
\citealt{Pakmor2014,Whittingham2021}), the circumgalactic medium of galaxies
(e.g. \citealt{Pakmor2020,vandeVoort2021}) or galaxy clusters (e.g.
\citealt{Dolag1999,Vazza2018,Quilis2020}).

An essential ingredient in building trust in the results of such computer simulations is verifying that the computer code used performs well
on a number of well-defined test problems. Many such tests are known from standard MHD (e.g. \citealt{Stone2008}) but very few have been developed specifically for comoving MHD (see Appendix~\ref{sec:current-comoving tests} for a brief review of these).

Due to the similarities between standard and comoving MHD, cosmological
codes are therefore mainly verified using tests developed for standard MHD. This is normally done by simply setting $a=1$ and $\dot{a}=0$ in the comoving MHD codes. This could be problematic because it makes it possible for errors related to $a\neq1$ and $\dot{a}\neq0$
to go unnoticed.
Since such errors could have far-reaching consequences for the validity of simulation results, we advocate in this paper for introducing additional tests that specifically target the comoving MHD equations.

We argue that the test cases should ideally have analytic solutions or, when this is not possible, have high-precision reference solutions which can be universally agreed upon. In addition, it is useful if the tests are computationally cheap since this encourages frequent testing.
Hydromagnetic waves in comoving coordinates fulfill these requirements because \emph{i}) analytic solutions can be derived and \emph{ii}) wave motion can be simulated in one spatial dimension (1D) such that tests can be done
within a few minutes on most personal computers. The purpose of this
paper is therefore to provide analytic reference solutions for hydromagnetic waves and to give practical examples of how they can be used to test comoving MHD implementations.

The rest of the paper is divided as follows: we first introduce the equations
of ideal MHD and their comoving counterpart in Section~\ref{sec:equations}.
We then derive analytic wave solutions in Einstein-de-Sitter (EdS) cosmology
in Section~\ref{sec:analytic-solutions}. In particular, Section~\ref{sec:analytic-alfven-wave}
contains the comoving Alfvén wave derivation while Section~\ref{sec:analytic-magnetosonic-wave} contains the derivation of analytic
solutions for comoving magnetosonic waves modified by self-gravity.
Moving beyond EdS, we also provide a brief description of a procedure for numerically solving the linearized equations for a $\Lambda$CDM cosmology in Section~\ref{sec:numeric-scipy-solutions}. These analytic and numeric reference solutions are used to test the comoving
MHD code \textsc{arepo} in Section~\ref{sec:arepo-examples}. The test section
is divided into subsections describing Alfvén wave tests (Section~\ref{sec:alfven-sim}) and compressible wave tests (Section~\ref{sec:compressible-simulations}). These test cases consider both standing and traveling waves as well as their convergence properties. We discuss the effects of self-gravity and magnetic fields on the compressible waves and how changing the adiabatic index leads to interesting
behaviour. We conclude the paper by discussing the merits of code testing in Section~\ref{sec:discussion}. Finally, the paper also contains three appendices which provide brief outlines of the code tests
currently in use for comoving MHD (Appendix~\ref{sec:current-comoving tests}), the transformation from standard MHD to comoving MHD (Appendix~\ref{app:transformation-of-equations}) and additional details about the analytic solutions (Appendix~\ref{app:analytic-details}).

\section{Comoving MHD equations with self-gravity}
\label{sec:equations}

We start by introducing the equations of standard MHD
and their comoving counterpart.
We take $\vec{r}$ to be the spatial coordinate
in a fixed coordinate system and $\vec{x}$ to be the
comoving coordinate. These are related by $\vec{r} = a \vec{x}$ where
$a$ is the time-dependent cosmological scale factor
which evolves according to the Friedmann equation
\be
    \dot{a} = a H_0 \left(\f{\Omega_{\mathrm{m},0}}{a^3}
    + \f{\Omega_{\mathrm{r},0}}{a^4}
    +
    \Omega_{\Lambda,0}
    \right)^{1/2}
    \label{eq:FLRW-simple} \ ,
\en
where $H_0$ is the Hubble parameter and
$\Omega_{\mathrm{m},0}$, $\Omega_{\mathrm{r},0}$, and $\Omega_{\Lambda,0}$
are the $z=0$ values of the cosmological parameters for total density
(baryonic plus dark matter), radiation and dark energy, respectively.
Cosmological MHD simulations generally start at redshifts low enough that the radiation density can effectively be ignored. We therefore set $\Omega_{\mathrm{r},0}=0$ throughout the paper.

The standard (i.e. non-comoving) equations of ideal MHD are
the mass continuity, momentum, induction
and entropy equations. In Heaviside-Lorentz units, they are given by (e.g. \citealt{Freidberg2014})
\be
    \der{\ln \rho}{t} = -\delr \bcdot \vec{\varv} \ ,
    \label{eq:rho}
\en
\be
    \rho \der{\vec{\varv}}{t} =
    - \delr p
    - \delr \bcdot \left(\f{B^2}{2}\mathbf{1} -
    \vec{B}\vec{B} \right) - \rho \delr \Phi\ ,
    \label{eq:mom}
\en
\be
    \pder{\vec{B}}{t} = \delr \btimes \left(\vec{\varv} \btimes \vec{B}\right)
    \label{eq:ind}\ ,
\en
\be
    \f{p}{\gamma -1}\der{\ln \left(p \rho^{-\gamma}\right)}{t} =
    0 \ ,
    \label{eq:ent}
\en
where $d/dt = \partial/\partial t + \vec{\varv} \bcdot \delr$ is the convective
derivative with $\partial/\partial t$ taken at fixed position $\vec{r}$,
$\rho$ is the gas density, $\vec{\varv}$ is the gas
velocity, $p$ is the thermal pressure, $\vec{B}$ is the magnetic field vector
with strength $B$, $\mathbf{1}$ is the unit tensor, $\vec{B}\vec{B}$ is a
dyadic product, $\Phi$ is the gravitational potential and $\gamma$ is the adiabatic index. We also note that the entropy equation can equivalently be written as
\be
    \der{\ln \varepsilon}{t} = -\gamma \delr \bcdot \vec{\varv} \ ,
    \label{eq:internal-energy}
\en
where $\varepsilon=p/(\gamma-1)$ is the internal energy.

The comoving MHD equations can be derived by transforming from the $\vec{r}$
coordinate system to the $\vec{x}$ coordinate system and including a cosmological source term in Poisson's equation for the gravitational potential. We provide a brief outline of a derivation in Appendix~\ref{app:transformation-of-equations}.\footnote{The
standard reference for the derivation of comoving
\emph{hydrodynamics} is section II of \citet{Peebles1980}.
Comoving MHD papers normally state the equations of comoving MHD
assuming $\gamma=5/3$ (e.g. \citealt{Pakmor2013,Bryan2014,Weinberger2020}). This assumption is relaxed in
our paper and we therefore include a brief outline of a derivation of comoving MHD for a general value of $\gamma$.
}
The result is
\be
    \der{\ln \rhoc}{t} = - \f{1}{a}\delx \bcdot \vec{u} \ ,
    \label{eq:rho-comoving}
\en
\begin{multline}
    a\rhoc \der{\vec{u}}{t} =
    - \delx \pc
    - \f{1}{a}\delx\bcdot\left(\f{\Bc^2}{2}\mathbf{1}
    - \vec{B}_\mathrm{c} \vec{B}_\mathrm{c}\right)
    - \rhoc\delx \delta \Phi
    - \rhoc \dot{a} \vec{u}\,,
    \label{eq:mom-comoving}
\end{multline}
\be
    \pder{\vec{B}_\mathrm{c}}{t}
    =
    \f{1}{a} \delx \times \left(\vec{u}\times \vec{B}_\mathrm{c} \right) \ ,
    \label{eq:Bc-comoving}
\en
\be
    \der{\epsc}{t} =
    -3\f{\dot{a}}{a}\left(\gamma-1\right)\epsc
    - \gamma \epsc
    \f{1}{a}\delx \bcdot \vec{u} \ ,
    \label{eq:epcs-comoving}
\en
and
\be
    \delx^2 \delta \Phi = \f{4 \upi G}{a} \delta \rho_{\mathrm{tot, c}} \ ,
    \label{eq:poisson-comoving}
\en
where $d/dt = \partial/\partial t + a^{-1} \vec{u} \bcdot \delx $ is the
convective derivative with $\partial/\partial t$ taken at fixed position $\vec{x}$, $\rhoc=\rho a^3$ is the comoving gas density, $\vec{u}$
is the peculiar velocity, $\vec{B}_\mathrm{c} = \vec{B}a^2$ is the comoving
magnetic field\footnote{It should be mentioned that there are other ways to define the comoving magnetic field, e.g., \citet{Li2008,Collins2010} take
$\vec{B}_\mathrm{c} = \vec{B}a^{3/2}$ while \citet{Rieder2017} use the super-comoving variables of \citet{Martel1998}. The definition used in the present paper, $\vec{B}_\mathrm{c} = \vec{B}a^2$, coincides with the one used in \textsc{arepo} \citep{Pakmor2013} and the more recent version of \textsc{enzo} \citep{Bryan2014}.} and $\epsc=\varepsilon a^3$ is the comoving internal energy. The density sourcing the Poisson equation
is $\delta \rho_{\mathrm{tot, c}} = \left(\rho_{\mathrm{tot, c}} - \bar{\rho}_{\mathrm{tot, c}}\right)$ where $\rho_\mathrm{tot, c}$ is the comoving total density (baryons, dark matter etc) and $\bar{\rho}_\mathrm{tot, c}$ is its mean (see details in Appendix~\ref{app:transformation-of-equations}).

\section{Analytic solutions}
\label{sec:analytic-solutions}

We derive analytic solutions for comoving MHD waves
by using linear perturbation theory. Our derivation is related to the extensive literature on cosmological perturbation theory which we discuss below.

The compressible modes in a hydrodynamic fluid coupled to self-gravity is
a well-known problem which is treated in most textbooks on cosmology (e.g. chapter 16 in \citealt{Peebles1980}). Most textbooks, e.g. \citet{Cimatti2019}, have a focus on scales that are dominated by gravitational instability. They therefore assume that the thermal pressure gradient can be neglected which has the advantage that it significantly simplifies the problem. In order to study
stable hydrodynamic waves modified by gravity it is however essential that the pressure gradient is retained. This is done in the book by \citealt{Weinberg1972} which discusses the Bessel function solution that then arises.

Moving beyond hydrodynamics is motivated by the belief that the early Universe contained primordial magnetic fields
\citep{Grasso2001,Durrer2013,Subramanian2016}.
The extension to MHD was pioneered by \citet{Wasserman1978} who considered
magnetic fields in the limit of zero thermal pressure gradient.
It was shown that magnetic fields can create density perturbations which could seed gravitational instability with consequences for large-scale structure formation.
This idea has since been further developed by e.g. \citet{Kim1996}, \citet{Tsagas2000}, \citet{Gopal2003}, and \citet{Shaw2012}.

Hydromagnetic waves in the early universe have also been extensively studied in the context of the cosmic microwave background (CMB, \citealt{PlanckCollaboration2016,PlanckCollaboration2020}). Such studies generally include a variety of physical effects beyond ideal MHD which are believed to be important prior to recombination ($z\sim 1100$).
For instance, \citet{Jedamzik1998} and \citet{Subramanian1998} considered neutrino and photon damping of Alfvén and magnetosonic waves. It is also common to consider a spectrum of waves since this makes a connection to observations \citep{Subramanian2006,Shaw2012}.

Other relevant work includes \citet{Holcomb1989}
who studied waves in a radiation-dominated cosmology in the ultra-relativistic limit and a follow-up work which considered a post-recombination plasma in a matter-dominated universe \citep{Holcomb1990}. These works were extended by \citet{Sil1996} who considered
a general $\gamma$ and a
cosmology parametrized by $a\propto t^n$ where $n$ is a free parameter (e.g. $n=2/3$ for a matter-dominated universe and
$n=1/2$ for a radiation-dominated universe).

The by far most comprehensive study of MHD waves in an expanding universe was given by \citet{Gailis1994,Gailis1995}
who derived the MHD eigenmodes for a general orientation of the magnetic field and wave vector. Given the breadth of their analysis, we find it likely that some of our analytic results can be shown to be special cases of their non-relativistic, matter-dominated solutions.
However, we note that \citet{Gailis1995} did not include self-gravity and that they focused on $\gamma=4/3$ and
$5/3$.\footnote{\citet{Gailis1995} assume that the temperature decreases as $a^{-1}$ pre-recombination and $a^{-2}$ post-recombination. This corresponds to $\gamma=4/3$ and $\gamma=5/3$, respectively.}
This is in contrast to our magnesonic wave
solution which considers a simple geometry but includes self-gravity and allows for a general adiabatic index.

\subsection{Spatially uniform and motionless background}
\label{sec:background}
Linear perturbation theory requires a background on which to perturb. We detail this background here.
We consider a universe which has no
peculiar motions and is spatially uniform (i.e. $\vec{u}$ and all gradients of the background are zero). It can be seen that the right-hand sides (RHSs) of equations~\eqref{eq:rho-comoving}--\eqref{eq:epcs-comoving} are all zero for such a background when $\gamma=1$.
The (well-known) consequence is that the background stays constant in time for such isothermal systems.
For non-isothermal systems, e.g. adiabatic with $\gamma=5/3$, the internal energy decays during cosmic expansion. This can be seen
directly from the non-zero RHS of equation~\eqref{eq:epcs-comoving}. The equation can be integrated to reveal that the comoving pressure,
$\pc=\epsc (\gamma-1)$ evolves as
\be
    \pc(a)
    = \pc(a_\mathrm{i})\,\left(\f{a}{a_\mathrm{i}}\right)^{-3(\gamma-1)}
    =
    \f{p_{\mathrm{c},0}}{\,a^{3(\gamma-1)}}\
\en
where $a_\mathrm{i}$ is the initial scale factor,
$\pc(a_\mathrm{i})$ is the comoving pressure at $a_\mathrm{i}$ and $p_{\mathrm{c},0}$ is the comoving pressure at $z=0$ (here the redshift, $z$, is related to the scale factor by $a = 1/(z+1)$ with $a=1$ at $z=0$).

\subsection{Characteristic wave speed and frequency definitions}
\label{sec:parameter-definitions}

We define the adiabatic sound speed $\cad=\sqrt{\gamma p/\rho}$ and the Alfv\'{e}n speed
$\va = B/\sqrt{\rho}$. With $B=\Bc a^{-2}$, $\rho=\rhoc a^{-3}$,
$p=\pc a^{-3}$ and $\pc=p_{\mathrm{c},0}a^{-3(\gamma-1)}$
their dependencies on redshift are given by
\be
    \cad
    \equiv
    \sqrt{\frac{\gamma p}{\rho}}
    =
    \sqrt{\frac{\gamma p_{\mathrm{c},0}}{\rhoc}} a^{-3(\gamma-1)/2} \ ,
    \label{eq:cad-definition}
\en
and
\be
    \va \equiv \f{B}{\sqrt{\rho}}
    =
    \f{\Bc}{\sqrt{\rhoc}} a^{-1/2} \ ,
    \label{eq:va-definition}
\en
respectively. We note that the sound speed depends on redshift when $\gamma\neq1$ but that it is independent
of redshift for an isothermal equation of state with $\gamma=1$.
The Alfv\'{e}n speed always depends on redshift. We also define a characteristic velocity
\be
    \vg
    = \f{\sqrt{4 \upi G \rhoc}}{k} a^{-1/2} \ ,
    \label{eq:vg-definition}
\en
which is associated with the self-gravity of the gas (here $k$ is the comoving wavenumber of the wave). We observe that $\vg$ and $\va$ have the same redshift dependence.

We have found it useful to introduce $\cado$, $\vao$ and $\vgo$ for the values of $\cad$, $\va$ and $\vg$ at $z=0$ ($a=1$). These are given by
\be
    \cado \equiv \sqrt{
          \frac{\gamma p_{\mathrm{c},0}}
           {\rhoc}
           } \ ,
    \quad
    \vao \equiv
    \f{\Bc}{\sqrt{\rhoc}} \ ,
    \quad
    \vgo \equiv \f{\sqrt{4 \upi G \rhoc}}{k} \ .
    \label{eq:V-definitions}
\en
Finally, we define corresponding dimensionless
frequencies
\be
    \Oms \equiv \f{k\cado}{H_0} \ ,
    \quad
    \Oma \equiv \f{k\vao}{H_0} \ ,
    \quad
    \Omg \equiv \f{k\vgo}{H_0} \ ,
    \label{eq:Omega-definitions}
\en
which will be used to further simplify the analytic results.

\subsection{Transformations and the EdS solution}

We have found it convenient to work with the scale factor, $a$,
instead of the time, $t$. The transformation rule between time derivatives and
scale factor derivatives is
\be
    \pder{f}{t} = \dot{a} \pder{f}{a} \ ,
    \label{eq:transform-1st-der}
\en
where $f$ is some function.
Our analytic solutions assume a flat, matter-dominated Einstein-de-Sitter (EdS) Universe in which $\Omega_{\mathrm{m},0}=1$ and $\Omega_{\mathrm{r},0}=\Omega_{\Lambda,0}=0$. It then follows from equation~\eqref{eq:FLRW-simple} that $\dot{a}=H_0/\sqrt{a}$ which allows for analytic progress.

\subsection{Comoving Alfvén wave}
\label{sec:analytic-alfven-wave}

We consider a linearly polarized comoving Alvén wave in EdS cosmology.
This type of wave is incompressible and neither equation~\eqref{eq:rho-comoving}
nor equation~\eqref{eq:epcs-comoving} enters the dynamics.
The linearized versions of the remaining equations, equations~\eqref{eq:mom-comoving} and \eqref{eq:Bc-comoving}, are given by
\be
    \pder{}{t}\f{\dBc}{\Bc} &=&  \f{\ui k}{a} \du \ , \\
    \pder{(a \du)}{t} &=& \ui k \f{\Bc^2}{a\rhoc} \f{\dBc}{\Bc} \ ,
\en
where we have taken the mean field to be in the $x$-direction and $\du$ and $\dBc$ to be in the $y$-direction. We transform the time-derivatives to $a$-derivatives, substitute $\dot{a}=H_0/\sqrt{a}$ as appropriate for EdS cosmology (see equation~\ref{eq:FLRW-simple}), and obtain
\be
    \pder{}{a}\f{\dBc}{\Bc} = \f{\ui k}{H_0 a^{1/2}} \du
    \ ,
    \label{eq:Bc-alfven} \\
    \pder{(a\du)}{a} = \f{\ui k\vao^2}{H_0 a^{1/2}} \f{\dBc}{\Bc}
    \label{eq:a-du-alfven}
    \ .
\en
We solve this coupled set of equations by combining them
\be
    \pder{}{a} \left(a^{3/2} \pder{}{a}\f{\dBc}{\Bc}
    \right)
    = \f{\ui k}{H_0} \pder{(a\du)}{a}
    =
    -\f{k^2\vao^2}{H_0^2 a^{1/2}} \f{\dBc}{\Bc}
    \ ,
\en
moving everything to the left-hand side (LHS)
and dividing through by $a^{3/2}$. This yields the second-order ordinary differential equation (ODE)
for $\dBc/\Bc$
\be
    \pdder{}{a}\f{\dBc}{\Bc}
    +
    \f{3}{2a}
    \pder{}{a}\f{\dBc}{\Bc}
    +
    \f{\Oma^2}{a^2} \f{\dBc}{\Bc}
    = 0 \ ,
    \label{eq:alfven-ODE}
\en
where $\Oma=k\vao/H_0$ was defined in equation~\eqref{eq:Omega-definitions}.

\begin{figure}
\includegraphics[trim= 0 20 0 2]{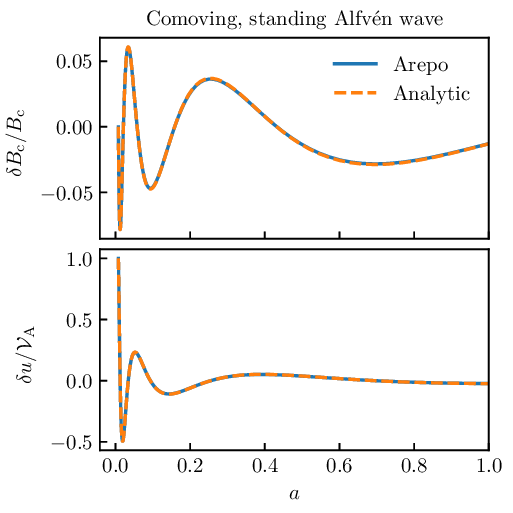}
\caption{Standing Alfvén wave in EdS cosmology (Sections~\ref{sec:analytic-alfven-wave} and \ref{sec:alfven-sim-standing}). The analytic solution is given by equations~\eqref{eq:analytic-alf-dBc-osc} and \eqref{eq:analytic-alf-du-osc} (Fourier amplitudes)
and in equations~\eqref{eq:analytic-alf-dBc-osc-real} and \eqref{eq:analytic-alf-du-osc-real} (real space with $\dBc(\ai)=0$).}
\label{fig:alfven-sim-standing-example}
\end{figure}

Equation~\eqref{eq:alfven-ODE} is an Euler differential equation and thus has a known analytic solution. We solve the indicial equation (see e.g. \citealt{asmar2010partial} pages A24-A25)
and find the solution
\be
    \f{\dBc}{\Bc} = a^{-1/4}\left(c_1 \ue^{\ui\kappa\ln a} + c_2 \ue^{-\ui\kappa\ln a}\right)
    \label{eq:alfven-Bc-gen-sol}
    \ ,
\en
where $c_1$ and $c_2$ are integration constants
and we have defined\footnote{
We note that the special case $\Oma=1/4$ (which gives $\kappa=0$) has a different solution. We provide this solution for completeness in Appendix~\ref{app:Oma=1/4}.
}

\be
    \kappa \equiv \sqrt{\Oma^2 - \f{1}{16}} \ .
    \label{eq:alfven-kappa}
\en
The solution for the peculiar velocity perturbation, $\du$, is found by differentiation using equation~\eqref{eq:Bc-alfven} and \eqref{eq:alfven-Bc-gen-sol}, this yields
\be
    \f{\du}{\vao} &=& -\f{\ui}{\Oma}\sqrt{a} \pder{}{a}\f{\dBc}{\Bc}
    = \nonumber \f{\ui}{\Oma}a^{-3/4} \times \\
    &&
    \left[c_1 \left(\f{1}{4} - \ui\kappa\right)\ue^{\ui\kappa\ln a}
    +
    c_2 \left(\f{1}{4} + \ui \kappa\right)\ue^{-\ui\kappa\ln a}\right]
    \label{eq:alfven-du-gen-sol}
     \ .
\en
The integration constants can be found using prescribed initial conditions
(i.e. the initial amplitudes in $\dBc$ and $\du$)
defined as $A_B \equiv \dBc(\ai)/\Bc$ and
$A_u \equiv \du(\ai)/\vao$. This is done by solving two equations for two unknowns which gives the result
\be
    \f{\dBc}{\Bc}
    =
    \left(\f{a}{\ai}\right)^{-1/4}
    \left[
    A_B \cos(\psi)
    + \f{A_B + 4\ui\Oma \sqrt{\ai}A_u}{4\kappa}
    \sin(\psi)
    \right]\ ,
    \label{eq:analytic-alf-dBc-osc}
    \\
    \f{\du}{\vao} =
    \left(\f{a}{\ai}\right)^{-3/4}
    \left[
    A_u \cos\left(\psi\right)
    -
    \f{A_u - 4 \ui \Oma \ai^{-1/2} A_B}{4\kappa}
    \sin\left(\psi\right)
    \right] \ ,
    \label{eq:analytic-alf-du-osc}
\en
where we have defined
\be
    \psi = \kappa \ln \left({a}/{\ai}\right) \ .
    \label{eq:psi-alfven}
\en
We show the solution for $A_B=0$, $A_u = 1$ in Fig.~\ref{fig:alfven-sim-standing-example}.

The solution for a comoving Alfvén wave in
EdS cosmology has many interesting features compared to the standard
Alfvén wave. We make the following observations:
The first, and perhaps rather obvious, observation, is that the
comoving Alfvén wave decays in amplitude during cosmic expansion.
Next, we observe that the Alfvén wave solution is periodic in $\ln a$, a feature also found for a shear Alfvén wave \citep{Holcomb1990}.
In addition, the comoving Alfvén wave cannot travel at all if the background
magnetic field is too weak or the wavelength is too large \citep{Holcomb1990,Gailis1995}. Indeed, $\kappa$ becomes imaginary when
$\Oma < 1/4$, such that the solutions given in equations~\eqref{eq:analytic-alf-dBc-osc} and
\eqref{eq:analytic-alf-du-osc} turn into power law decays.

Finally, we can calculate the frequency of the comoving Alfvén wave.
Taking the time derivative of the phase $\psi$ (see equation~\ref{eq:psi-alfven}),
we find
\be
    \dot{\psi} =
    k \vao a^{-3/2} \sqrt{1 - \left(\f{H_0}{4k\vao}\right)^2} \ ,
    \label{eq:psi-dot-alfven}
\en
where we have used that $\dot{a}/a=H_0 a^{-3/2}$ for EdS cosmology.
We compare this to the standard Alfvén wave frequency
$\omega_\mathrm{A}= k' \va$
where $k'$ is the physical wavenumber and $\va$ is the Alfvén speed.
Since the wavenumber and Alfvén speed decrease as $k'=k/a$
and $\va = \vao a^{-1/2}$ during cosmic expansion, the standard
MHD theory predicts that the Alfvén frequency decreases
as $\omega_\mathrm{A}\propto a^{-3/2}$.
This  $a^{-3/2}$ scaling is indeed seen in equation~\eqref{eq:psi-dot-alfven}.
The additional square root dependence shows that the comoving
Alfvén waves are dispersive, i.e., that $\dot{\psi}/k$
depends on $k$. This modification of the wave frequency,
and thus the dispersion effect, is most significant for
waves close to the damping scale given by $\Oma\leq1/4$.
Indeed, this dispersion feature is closely related to the
damping of the wave, with the frequency of the wave
going to zero as the damping becomes maximal.
While the damping is here due to cosmic expansion,
a similar relation between damping and dispersion
is generally found for dissipative waves
(see e.g. \citealt{Berlok2020} for waves damped by Braginskii viscosity).

The solution given by equations~\eqref{eq:analytic-alf-dBc-osc} and \eqref{eq:analytic-alf-du-osc} works well for comparing with simulations
where arbitrary values of $A_B$ and $A_u$ are initialized.
However, a wave eigenmode has a specific
value of $A_B/A_u$. We require this eigenmode as we also wish to initialize a
traveling wave in the code testing section. We therefore determine the eigenmode and thus $A_B/A_u$ as follows:
the plus and minus in $\ue^{\pm \ui \kappa \ln a}$ in equation~\eqref{eq:alfven-Bc-gen-sol}
correspond to waves traveling in opposite directions. We can without loss of generality focus on the plus solution, for which we observe that both $\dBc/\Bc a^{1/4}$ and $\du/\vao a^{3/4}$ are proportional to $\ue^{\ui \kappa \ln a}$. Using this information, we can write equations~\eqref{eq:Bc-alfven} and \eqref{eq:a-du-alfven} as
an eigenvalue problem
\be
    \left(
    \begin{matrix}
    -\ui/4 - \kappa & \Oma \\
    \Oma & \ui/4 - \kappa
    \end{matrix}
    \right)
    \left(
    \begin{matrix}
    \dBc/\Bc a^{1/4} \\
    \du/\vao a^{3/4}
    \end{matrix}
    \right)
    =
    \left(
    \begin{matrix}
    0 \\
    0
    \end{matrix}
    \right) \ ,
\en
where $\kappa$ is the eigenvalue.
Solving this eigenvalue problem, we find the eigenmode solutions
\be
    \f{\dBc}{\Bc} &=& C \f{\pm 4\kappa - \ui}{4\Oma} a^{-1/4} \ue^{\pm \ui\kappa \ln a} \ ,
    \label{eq:alfven-dBc-eigenmode} \\
    \f{\du}{\vao} &=& C a^{-3/4} \ue^{\pm \ui\kappa \ln a} \ ,
    \label{eq:alfven-du-eigenmode}
\en
where $C$ is the overall mode amplitude. The eigenmode thus has $A_B/A_u=\sqrt{\ai}(\pm4\kappa-\ui)/4\Oma$.

\subsection{Comoving magnetosonic wave with self-gravity}
\label{sec:analytic-magnetosonic-wave}

We derive analytic solutions for comoving magnetosonic waves modified by self-gravity in EdS cosmology. We
take the background magnetic field and its perturbation, $\dBc$, to be in the $z$-direction and the wave vector and the velocity perturbation, $\du$, to be in the $x$-direction.
The linearized gas density equation, equation~\eqref{eq:rho-comoving}, is
\be
    \pder{}{t}\f{\drhoc}{\rhoc} = - \f{\ui k}{a}\du\ ,
\en
the linearized momentum equation, equation~\eqref{eq:mom-comoving}, is
\be
    \pder{(a \du)}{t} = -\ui k \left(\frac{\dpc}{\rhoc}
    -\f{4 \upi G \rhoc}{k^2 a}\f{\drhoc}{\rhoc}
    +\f{\Bc^2}{a\rhoc} \f{\dBc}{\Bc}\right)\ ,
    \label{eq:a-du-dt-comp}
\en
the linearized induction equation, equation~\eqref{eq:Bc-comoving}, is
\be
    \pder{}{t}\f{\dBc}{\Bc} = - \f{\ui k}{a} \du = \pder{}{t}\f{\drhoc}{\rhoc}\ ,
    \label{eq:Bc-dt-comp}
\en
and the linearized internal energy equation,  equation~\eqref{eq:epcs-comoving}, is
\be
    \pder{\dpc}{t} =
    -3\f{\dot{a}}{a}\left(\gamma-1\right)\dpc
    - \f{\gamma p_{\mathrm{c},0}}{\,a^{3(\gamma-1)}} \f{\ui k}{a} \du
    \ .
    \label{eq:dpc-comoving}
\en
We make the simplifying assumption for the initial condition that $\dBc/\Bc = \drhoc/\rhoc$ (which, given
equation~\ref{eq:Bc-dt-comp}, means that $\dBc/\Bc = \drhoc/\rhoc$ at all times).
The linearized momentum equation, equation~\eqref{eq:a-du-dt-comp}, can then be written as
\be
    \pder{(a \du)}{t} = -\ui k \frac{\dpc}{\rhoc}
    -\f{\ui k}{a} \left(\f{\Bc^2}{\rhoc}-\f{4 \upi G \rhoc}{k^2}\right)
    \f{\drhoc}{\rhoc}\ .
\en
We transform the $t$-derivatives to $a$-derivatives and obtain
\be
    \pder{}{a}\f{\drhoc}{\rhoc} = - \f{\ui k}{a\dot{a}}\du\ ,
    \label{eq:drho-da-comp}
\en
\be
    \pder{(a\du)}{a} = - \f{\ui k}{\dot{a}} \frac{\dpc}{\rhoc}
    -\f{\ui k}{a\dot{a}} \left(\f{\Bc^2}{\rhoc}-\f{4 \upi G \rhoc}{k^2}\right)
    \f{\drhoc}{\rhoc}\ .
\en
\be
    \pder{\dpc}{a} =
    -3\f{1}{a}\left(\gamma-1\right)\dpc
    - \f{\gamma p_{\mathrm{c},0}}{\,a^{3(\gamma-1)}} \f{\ui k}{a\dot{a}} \du \ .
    \label{eq:dpc-da-comp}
\en
We combine the continuity equation with the internal energy equation
(equations~\ref{eq:drho-da-comp} and \ref{eq:dpc-da-comp})
and find
\be
    \pder{\dpc}{a} =
    -3\f{1}{a}\left(\gamma-1\right)\dpc
    + \f{\gamma p_{\mathrm{c},0}}{\rhoc\,a^{3(\gamma-1)}} \pder{\drhoc}{a} \ .
\en
We rewrite this equation as
\be
    \pder{}{a}\left(a^{3(\gamma-1)} \dpc\right) =
    \cado^2 \pder{\drhoc}{a} \ ,
\en
which can immediately be integrated to obtain
\be
    \dpc = \f{\cado^2}{a^{3(\gamma-1)}} \drhoc = \cad^2 \drhoc \ .
    \label{eq:dp-drho-relation}
\en
Equation~\eqref{eq:dp-drho-relation} is the standard result for the relation between pressure and density perturbations in a sound wave. However, when $\gamma\neq1$ the sound speed depends on redshift due to the decrease in temperature that takes place during cosmic expansion.

The simplifications outlined above reduce the system to two coupled, linear ODEs with non-constant coefficients.
Inserting $\dot{a}=H_0/\sqrt{a}$ for EdS cosmology
and our definitions for $\cado$, $\vao$ and $\vgo$
these equations are given by
\begin{align}
    \pder{}{a}\f{\drhoc}{\rhoc} &= - \f{\ui k}{H_0 \sqrt{a}}\du\ ,
    \label{eq:drhoc-da-comp-final}
    \\
    \pder{(a\du)}{a} &=
    -\f{\ui k\sqrt{a}}{H_0} \left(\f{\cado^2}{a^{3(\gamma-1)}} + \f{\vao^2-\vgo^2}{a}\right)
    \f{\drhoc}{\rhoc}
    \label{eq:adu-da-comp-final}
    \ .
\end{align}
We next combine equations~\eqref{eq:drhoc-da-comp-final} and \eqref{eq:adu-da-comp-final} to obtain
\begin{multline}
   \pder{}{a} \left(a^{3/2}\pder{}{a}\f{\drhoc}{\rhoc}
   \right) = - \f{\ui k}{H_0} \pder{(a\du)}{a}
   =  \\
   -\f{k^2 a^{1/2}}{H_0^2} \left(\f{\cado^2}{a^{3(\gamma-1)}} + \f{\vao^2-\vgo^2}{a}\right)
    \f{\drhoc}{\rhoc}
   \ ,
\end{multline}
which we simplify by moving everything to the LHS, dividing through by $a^{3/2}$
and introducing the dimensionless frequencies defined in equation~\eqref{eq:Omega-definitions}.
This procedure yields a second order, linear, non-constant coefficient ODE for $\drhoc/\rhoc$:
\begin{multline}
   \pdder{}{a}\f{\drhoc}{\rhoc}
   +
   \f{3}{2a}\pder{}{a}\f{\drhoc}{\rhoc}
   +
   \left(\f{\Oms^2}{a^{3\gamma-2}} + \f{\Oma^2-\Omg^2}{a^2}\right)
    \f{\drhoc}{\rhoc}
    = 0
    \label{eq:rhoc-ODE-comp}
   \ .
\end{multline}
We solve equation~\eqref{eq:rhoc-ODE-comp} in the following two subsections. We start with the special case $\gamma=4/3$ in Section~\ref{sec:comp-gamma-4/3} (which is the adiabatic index for a relativistic gas) and then proceed with a more general $\gamma\neq4/3$ in Section~\ref{sec:comp-gamma-not-4/3}. The $\gamma=4/3$ solution has the advantage that it can be written in terms of elementary functions while the $\gamma\neq4/3$
solution has the advantage that it includes commonly used values of $\gamma$ (in particular $\gamma=1$ and $5/3$).

Before proceeding with these detailed derivations, we can already now comment
on the criterion for gravitational instability of
the wave.\footnote{Here we discuss the criterion for gravitational instability
of a single wave with a fixed comoving wavelength. The normally considered, and astrophysically
more relevant, discussion is how the physical length
scale above which gravitational instability sets in
(i.e. the Jeans length) evolves as function of scale
factor (e.g. \citealt{Barkana2001}).}
A necessary (but not always sufficient, see
Section~\ref{sec:iso-wave-sim}) condition for gravitational instability is
that the restoring force of the wave disappears,
i.e., if the terms inside the parenthesis in equation~\eqref{eq:adu-da-comp-final}
become negative. This criterion can be written\footnote{While equation~\eqref{eq:adu-da-comp-final}
has already specialized to EdS cosmology the discussion pertains to $\Lambda$CDM as well.
}
\be
    a^{4-3\gamma} < \f{\vgo^2-\vao^2}{\cado^2} \ .
    \label{eq:Jeans-criterion}
\en
For $\gamma=4/3$ the wave is either gravitationally stable at all times or
gravitationally unstable at all times. This behavior happens because $\cad$,
$\va$ and $\vg$ all have the same $a$-dependence when $\gamma=4/3$.
For $\gamma<4/3$, $\va$ and $\vg$ decay faster than $\cad$. This means that
the wave can be gravitationally unstable at early times and then transition to
stability at later times. Linear solutions can
describe this behavior as long as the time period for growth is short
enough that the wave amplitude remains
small. We show an example of this with $\gamma=1$ in Section~\ref{sec:iso-wave-sim}.
For $\gamma>4/3$, the restoring force provided by thermal pressure decays faster than the
gravitational (and magnetic) forces. In such cases, the wave will eventually
fall prey to gravitational instability unless $\vao>\vgo$.
We show an example of this behavior for $\gamma=5/3$ in Section~\ref{sec:adi-magneto-sim-EdS}.

\subsubsection{Magnetosonic waves with self-gravity and $\gamma=4/3$}
\label{sec:comp-gamma-4/3}

\begin{figure*}
\includegraphics[trim= 0 25 0 15]{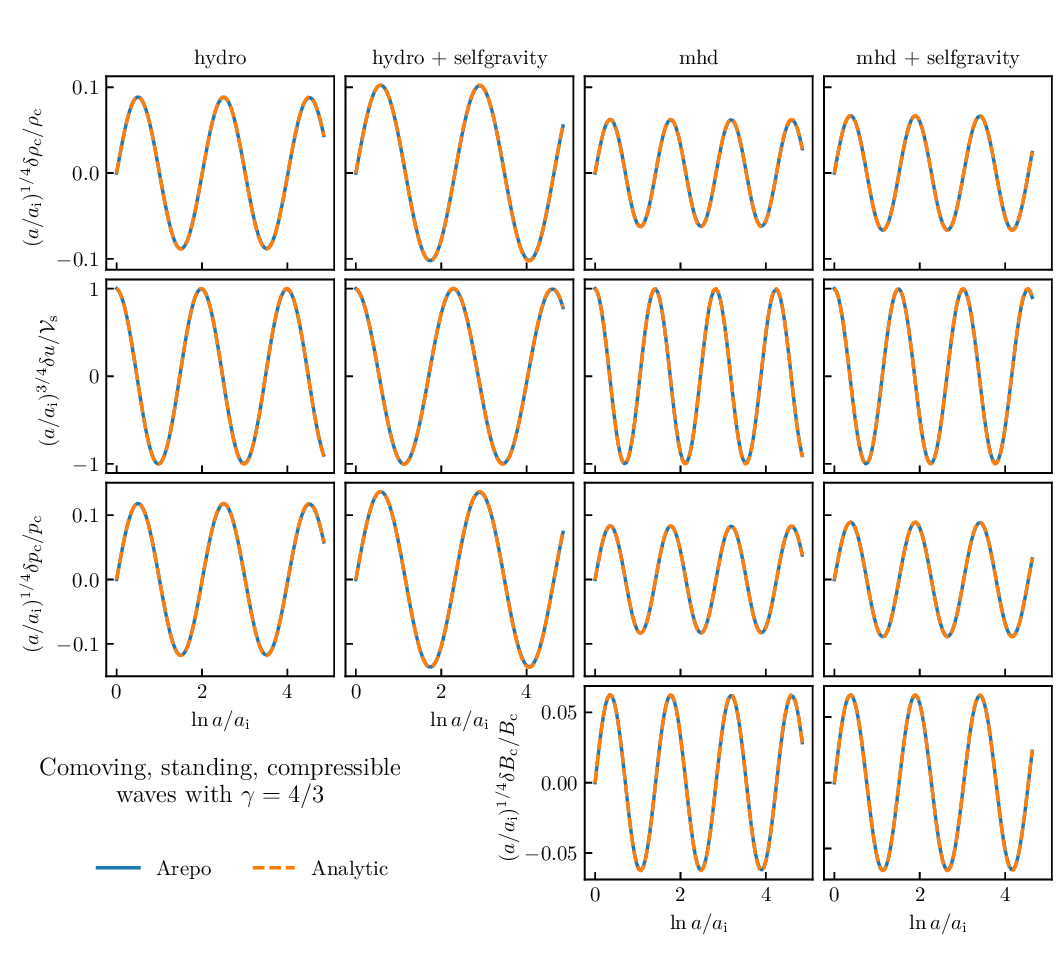}
\caption{Comparison between analytic theory and \textsc{arepo} simulations of
a compressible, standing perturbation in a gas with adiabatic index $\gamma=4/3$ residing in an EdS universe (Sections~\ref{sec:comp-gamma-4/3} and \ref{sec:compressible-gamma-4/3-sim-standing}).
We show hydrodynamics (first column), hydrodynamics plus self-gravity (second column), MHD (third column) and MHD plus self-gravity (last column).
The wave frequency decreases when self-gravity is included and increases
when magnetic fields are included.
The analytic solutions are given by equations~\eqref{eq:analytic-drho-standing-gamma-4/3} and
\eqref{eq:analytic-du-standing-gamma-4/3} (Fourier amplitudes) or equations~\eqref{eq:analytic-drho-standing-gamma-4/3-real} and
\eqref{eq:analytic-du-standing-gamma-4/3-real} (real space assuming $\drhoc(\ai)=0$).
}
\label{fig:compressible-gamma-4/3-standing}
\end{figure*}

We solve equation~\eqref{eq:rhoc-ODE-comp} for the special case $\gamma=4/3$. In this case, the equation reduces to
\be
    \pdder{}{a}\f{\drhoc}{\rhoc}
    +
    \f{3}{2a}
    \pder{}{a}\f{\drhoc}{\rhoc}
    +
    \f{\sigma^2}{a^2} \f{\drhoc}{\rhoc}
    = 0 \ ,
    \label{eq:ODE-gamma-4/3}
\en
where we have defined
\be
    \sigma \equiv \sqrt{\Oms^2 + \Oma^2 - \Omg^2}
    \label{eq:sigma-definition}
     = \f{k}{H_0} \sqrt{\cado^2 + \vao^2 - \vgo^2} \ .
\en
As for the comoving Alfvén wave, this ODE is an Euler equation. Solving the indicial equation, we find the solution for the density to be\footnote{As
for the Alfvén wave, we note that $\kappa=0$ requires special treatment. See Appendix~\ref{app:sigma=1/4} for the solution
for this particular value.}
\be
    \f{\drhoc}{\rhoc} = a^{-1/4}\left(c_1 \ue^{\ui\kappa\ln a} + c_2 \ue^{-\ui\kappa\ln a}\right)
    \ ,
    \label{eq:gen-solution-gamma-4/3}
\en
where
\be
    \kappa = \sqrt{\sigma^2 - \f{1}{16}} \ .
    \label{eq:kappa-gamma-4/3}
\en
The velocity solution is given by
\be
    \f{\du}{\cado} = \f{\ui \sqrt{a}}{\Oms} \pder{}{a}\f{\drhoc}{\rhoc} \ ,
\en
and the solution for initial condition $\drhoc(\ai)/\rhoc=A_\rho$ and $\du(\ai)/\cado=A_u$ is
\be
    \f{\drhoc}{\rhoc} =
    \left(\f{a}{\ai}\right)^{-1/4}
    \left[
    A_\rho \cos(\psi)
    + \f{A_\rho - 4\ui\Oms \sqrt{\ai}A_u}{4\kappa}
    \sin(\psi)
    \right]\ ,
    \label{eq:analytic-drho-standing-gamma-4/3}
\en
\begin{multline}
    \f{\du}{\cado} =
    \left(\f{a}{\ai}\right)^{-3/4}
    \left[
    A_u\cos\left(\psi\right)
    -
    \left(\f{A_u}{4\kappa}
    + \ui A_\rho \f{1 + 16 \kappa^2}{16\sqrt{\ai}\kappa \Oms}
    \right)
    \sin\left(\psi\right)
    \right] \ ,
    \label{eq:analytic-du-standing-gamma-4/3}
\end{multline}
where $\psi = \kappa \ln \left({a}/{\ai}\right)$.
This solution simplifies considerably when $A_\rho=0$. We show examples of
the solution with this initial condition in Fig.~\ref{fig:compressible-gamma-4/3-standing}.
The solutions all have $\Oms=\upi$ with $\Oma =\upi$ when magnetic fields are included and $\Omg = \upi/2$ when self-gravity is included.

\begin{figure*}
\includegraphics[trim= 0 20 0 10]{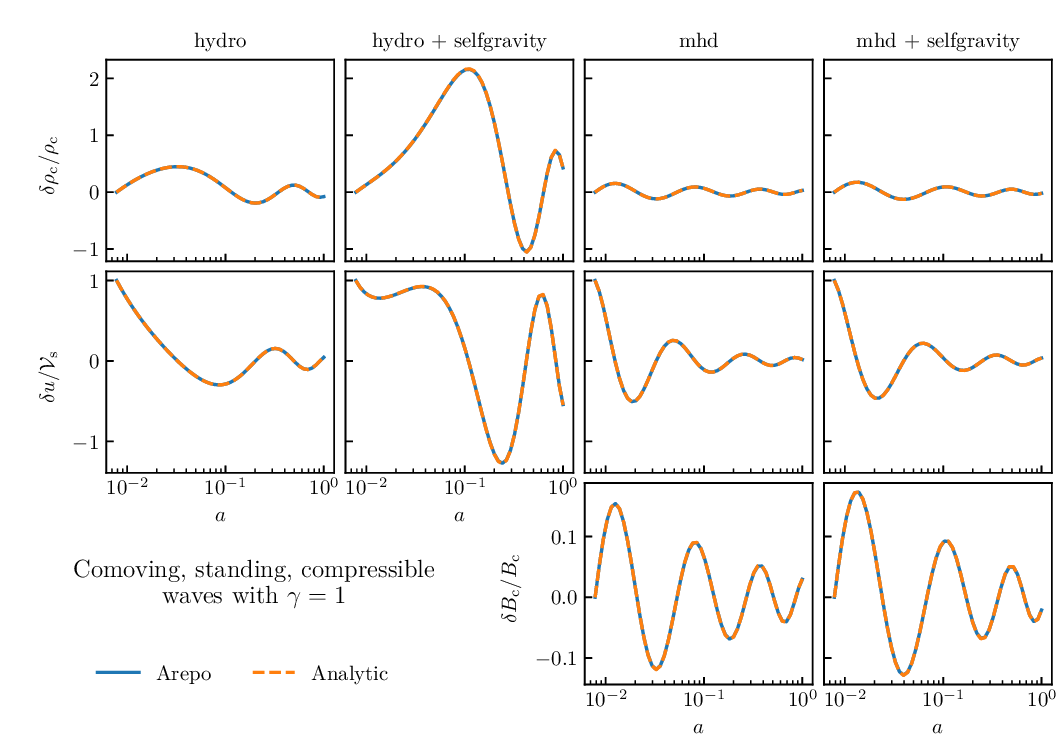}
\caption{Comparison between analytic theory and \textsc{arepo} simulation of
a compressible, standing perturbation in an isothermal gas with adiabatic index $\gamma=1$ residing in an EdS universe (Sections~\ref{sec:comp-gamma-not-4/3} and \ref{sec:iso-wave-sim}).
The hydrodynamic simulation with self-gravity has a larger amplitude than the others because the wave is gravitationally unstable for $a < 1/16$ (see equation~\ref{eq:Jeans-criterion}). The magnetic field
in the MHD simulation with self-gravity is strong enough to fully prevent gravitational instability but the wave frequency is still decreased compared to
the MHD simulation without self-gravity.
The analytic solutions are given by equations~\eqref{eq:drho-besselY-solution} and \eqref{eq:du-besselY-solution} (Fourier amplitudes). The hydrodynamic solution without self-gravity can be written in simple form, see equations~\eqref{eq:hydro-drhoc-gamma1} and \eqref{eq:hydro-du-gamma1} (real space assuming $\drhoc(\ai)=0$).}
\label{fig:compressible-gamma-1-EdS}
\end{figure*}

This solution for a comoving magnetosonic wave modified by self-gravity
in EdS cosmology has several interesting features. We
observe from equation~\eqref{eq:sigma-definition} that
$\sigma^2$ can be either positive or negative, depending on the relative
strengths of $\Oms$, $\Oma$ and $\Omg$ (or alternatively, $\cado$, $\vao$ and $\vgo$).

The solution consists of traveling (or oscillating) waves whose
amplitude decay with power law rates ($a^{-1/4}$ for $\drhoc/\rhoc$ and
$a^{-3/4}$ for $\du/\cado$) when $\sigma^2> 1/16$. As in fixed frame MHD, the magnetic field increases the frequency of the wave while self-gravity decreases the frequency of the wave (see equations~\ref{eq:sigma-definition} and \ref{eq:kappa-gamma-4/3}). The oscillatory
feature of the solution disappears when $0<\sigma^2<1/16$. In this case the perturbations instead suffer a power law decay
in a manner similar to the weak magnetic field solution found for comoving Alfvén
waves.

Finally, $\kappa$ becomes imaginary with $|\kappa| > 1/4$ when $\sigma^2<0$
and the amplitude of the perturbation therefore grows in time. The underlying physical mechanism is gravitational instability of the wave which is seen from equation~\eqref{eq:sigma-definition} to occur when $\vgo^2 > \cado^2 + \vao^2$. This criterion agrees with
equation~\eqref{eq:Jeans-criterion} which
was heuristically derived.

The compressible waves with $\gamma=4/3$ are dispersive. We show this by calculating the wave frequency by taking the time derivative of $\psi$, this yields
\be
    \dot{\psi} = k\left(\cado^2 + \vao^2 - \vgo^2\right)^{1/2}
    a^{-3/2} \sqrt{1 - \f{1}{16\sigma^2}} \ .
    \label{eq:dot-psi-gamma-4/3}
\en
Since $\sigma$ depends on $k$ and enters the square root in equation~\eqref{eq:dot-psi-gamma-4/3} the waves with $\gamma=4/3$ are dispersive, in particular for waves close to the damping scale.
A comparison between equation~\eqref{eq:dot-psi-gamma-4/3} and the fixed frame wave frequency (see e.g. \citealt{Pringle2014})
\be
    \omega = k'\left(\cad^2 + \va^2 - \vg^2\right)^{1/2}= k a^{-3/2}\left(\cado^2 + \vao^2 - \vgo^2\right)^{1/2} \ ,
    \label{eq:omega-compressible-standard}
\en
reveals that the $a^{-3/2}$ dependence given in equation~\eqref{eq:dot-psi-gamma-4/3} is as expected from the decrease in wavenumber, $k'=k/a$, and the $a^{-1/2}$ decay of $\cad$, $\va$ and $\vg$ that occurs during cosmic expansion when $\gamma=4/3$.

As for the Alfvén wave, we are also interested in
obtaining the eigenmode in order to be able to initialize
traveling waves. We use our knowledge of the solution (equations~\ref{eq:analytic-drho-standing-gamma-4/3} and \ref{eq:analytic-du-standing-gamma-4/3})
to write equations~\eqref{eq:drhoc-da-comp-final} and
\eqref{eq:adu-da-comp-final} as the eigenvalue problem
\be
    \left(
    \begin{matrix}
    -\ui/4 - \kappa & -\Oms \\
    -\sigma^2/\Oms & \ui/4 - \kappa
    \end{matrix}
    \right)
    \left(
    \begin{matrix}
    \drhoc/\rhoc a^{1/4} \\
    \du/\cado a^{3/4}
    \end{matrix}
    \right)
    =
    \left(
    \begin{matrix}
    0 \\
    0
    \end{matrix}
    \right) \ .
\en
Solving this eigenvalue problem, we find that the eigenvalues $\pm\kappa$ have eigenmodes given by
\be
    \f{\drhoc}{\rhoc} &=& -C \Oms \f{\pm 4\kappa - \ui }{4\sigma^2} a^{-1/4} \ue^{\pm \ui\kappa \ln a} \ ,
    \label{eq:magneto-drhoc-fourier-traveling-gamma-4/3}
    \\
    \f{\du}{\cado} &=& C a^{-3/4} \ue^{\pm \ui\kappa \ln a} \ ,
    \label{eq:magneto-du-fourier-traveling-gamma-4/3}
\en
where $C$ is the mode amplitude.

\subsubsection{Magnetosonic waves with self-gravity and $\gamma\neq4/3$}
\label{sec:comp-gamma-not-4/3}

\begin{figure*}
\includegraphics[trim= 0 20 0 10]{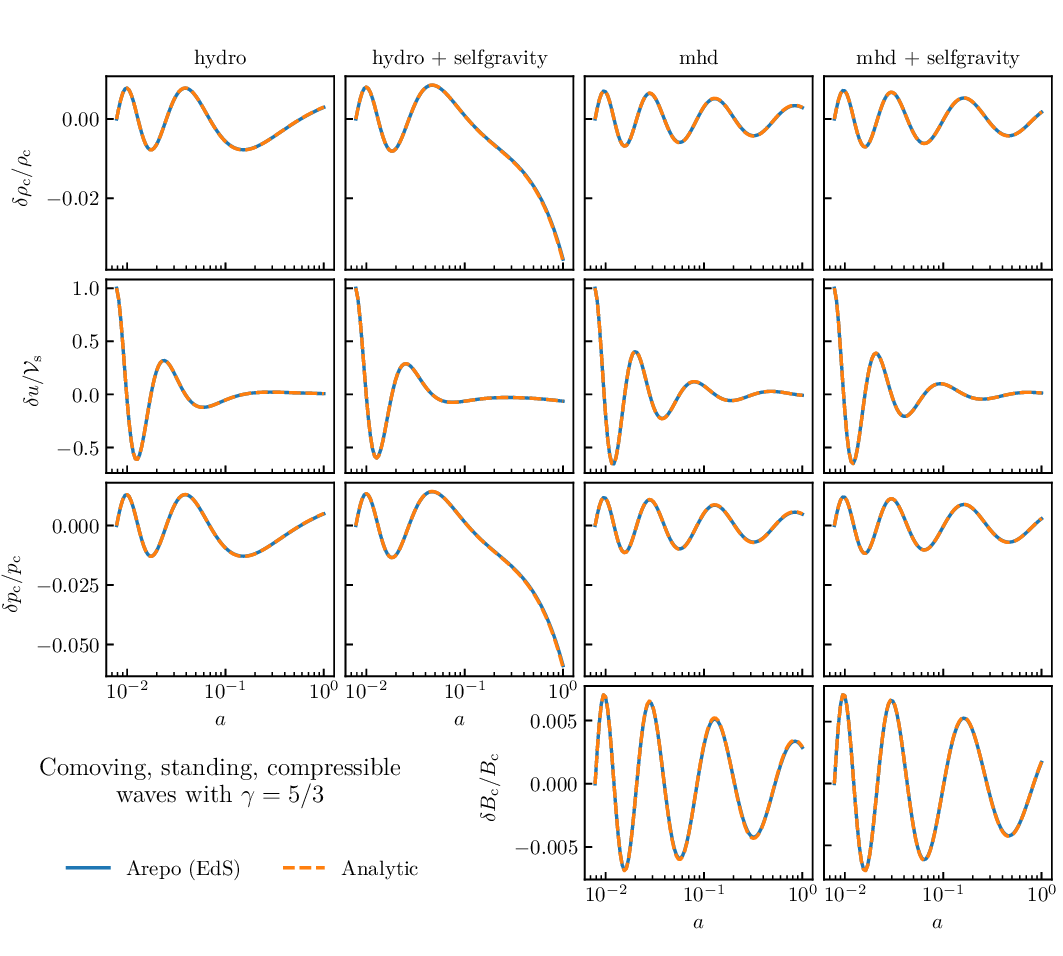}
\caption{Comparison between analytic theory and \textsc{arepo} simulation of
a compressible, standing perturbation in a gas with adiabatic index $\gamma=5/3$ residing in an EdS universe (Sections~\ref{sec:comp-gamma-not-4/3} and \ref{sec:adi-magneto-sim-EdS}).
We show hydrodynamics (first column), hydrodynamics plus self-gravity (second column), MHD (third column) and MHD plus self-gravity (last column).
The hydrodynamic simulation with self-gravity is seen to go gravitationally unstable towards the end of the simulation. This is in agreement with
equation~\eqref{eq:Jeans-criterion} which for our parameters predicts that this should occur when $a>4/25$. The magnetic field in the MHD simulation with
self-gravity is strong enough to prevent the gravitational instability.
The general analytic solutions are given by  equations~\eqref{eq:drho-besselY-solution} and \eqref{eq:du-besselY-solution}
(Fourier amplitudes) but
the hydrodynamic solution without self-gravity can be written in a much simpler form, see equations~\eqref{eq:drhoc-gamma5/3} and \eqref{eq:du-gamma5/3} (real space assuming $\drhoc(\ai)=0$).
}
\label{fig:compressible-gamma-5/3-EdS}
\end{figure*}

By comparing equation~\eqref{eq:rhoc-ODE-comp} with equation 6.80 in §104 in \citet{Bowman1958} we find that it is a transformed Bessel equation which has a
known analytic solution (see details in Appendix~\ref{app:magnetosonic-derivation-details}).
By introducing two additional parameters,
\be
    s \equiv \f{4-3\gamma}{2} \ ,
    \label{eq:s-parameter}
\en
\be
    \nu \equiv \f{
    \sqrt{1 - 16{\left(\Oma^2-\Omg^2\right)}
    }
    }{4|s|} \ ,
    \label{eq:nu-parameter}
\en
the solution for $\drhoc/\rhoc$ can be written as
\be
    \f{\drhoc}{\rhoc} = c_1 \mathcal{F}(a) + c_2 \mathcal{G}(a) \ ,
    \label{eq:drho-besselY-solution}
\en
where $c_1$ and $c_2$ are integrations constants and the functions $\mathcal{F}(a)$ and $\mathcal{G}(a)$ are given by\footnote{In the literature (e.g. \citealt{Holcomb1990,Gailis1995}), compressible wave solutions like the one presented here are often written in
terms of Hankel functions of the first and second kind, $H^{1}_\nu(z)$ and
$H^{2}_\nu(z)$.
Using the relations $H^{1}_\nu(z) = J_\nu(z) + \ui Y_\nu(z)$ and $H^{2}_\nu(z) = J_\nu(z) - \ui Y_\nu(z)$ our solution can be converted to this form.}
\be
    \mathcal{F}(a) = a^{-1/4} J_\nu\left(z\right) \ , \quad
    \mathcal{G}(a) = a^{-1/4} Y_\nu\left(z\right) \ .
    \label{eq:FandG}
\en
Here $J_\nu\left(z\right)$ and $Y_\nu\left(z\right)$
are Bessel functions of the first and second kind, respectively,
and we have introduced $z=\Oms a^s /|s|$ as a short-hand for their argument.\footnote{This Bessel function solution has divisions by zero for $s=0$ which occurs when $\gamma=4/3$. The solution for this specific value of $\gamma$ was given separately in section~\ref{sec:comp-gamma-4/3}.}

The solution for $\du$ is found by differentiation using equation~\eqref{eq:adu-da-comp-final} and is given
by
\be
    \f{\du}{\cado}
    =
    i \f{\sqrt{a}}{\Oms}
    \Big[ c_1 \mathcal{F}'(a) + c_2 \mathcal{G}'(a) \Big] \ .
    \label{eq:du-besselY-solution}
\en
The integration constants, $c_1$ and $c_2$,
 are determined from the initial conditions, $A_\rho =\drhoc(\ai)/\rhoc$ and $A_u = \du(\ai)/\cado$.
The details are given in Appendix~\ref{app:magnetosonic-derivation-details} with the results for $c_1$ and $c_2$ in equations~\eqref{eq:c1-magnetosonic-general-gamma} and
\eqref{eq:c2-magnetosonic-general-gamma}.

The solution simplifies considerably when initial conditions with $A_\rho=0$ are considered. In this case, the solution is given by
\be
    \f{\drhoc}{\rhoc}
    &=&
    \,\,\,\,A_u \ui \Oms \f{\upi \ai}{2 s}\Big[\mathcal{G}(\ai) \mathcal{F}(a)
    -
    \mathcal{F}(\ai) \mathcal{G}(a)
    \Big] \ ,
    \label{eq:drhoc-analytic-magnetosonic-only-Au}
    \\
    \f{\du}{\cado}
    &=&
    - A_u \sqrt{a} \f{\upi \ai}{2 s}\Big[\mathcal{G}(\ai) \mathcal{F}'(a)
    -
    \mathcal{F}(\ai) \mathcal{G}'(a)
    \Big]\ .
    \label{eq:du-analytic-magnetosonic-only-Au}
\en
The effects of magnetic field and self-gravity are contained in the $\nu$
parameter given in equation~\eqref{eq:nu-parameter}. This parameter is real if the magnetic field is weak but it becomes imaginary if the magnetic field strength is strong
enough to make the radicand negative, i.e., if $\Oma^2 > 1/16 + \Omg^2$. As
described in \citet{Matyshev2009}, Bessel functions of imaginary order appear
less frequently in scientific applications than the more standard Bessel
functions of real order. Perhaps for this reason, not all Bessel function
libraries allow for evaluation of non-real order functions (e.g. the
\textsc{scipy} special function library only supports real order at the time of
writing). We note that both \textsc{mathematica} and the Python library
\textsc{mpmath} \citep{mpmath} are able to evaluate Bessel functions for
imaginary $\nu$. We make our Python implementation of
equations~\eqref{eq:drho-besselY-solution} and \eqref{eq:du-besselY-solution}
available online in order to make the analytic solution more easily accessible.

We can find the analytic solution for comoving soundwaves
by ignoring the effects of magnetic field and self-gravity. The resulting expressions are given solely in terms of elementary functions.
We are particularly interested in isothermal ($\gamma=1$)
and adiabatic ($\gamma=5/3$) sound waves (sections~\ref{sec:iso-wave-sim} and \ref{sec:adi-magneto-sim-EdS}). Both these cases
have $|s|=1/2$ which gives $\nu=1/2$ when $\Oma=\Omg=0$. The analytic solutions given by equations~\eqref{eq:drhoc-analytic-magnetosonic-only-Au} and \eqref{eq:du-analytic-magnetosonic-only-Au} then simplify dramatically as the Bessel functions reduce as follows \citep{Abramowitz1972}
\be
    \f{J_{1/2}(z)}{\sin(z)} = -
    \f{Y_{1/2}(z)}{\cos(z)}
    =
    \sqrt{\frac{2}{\upi z}} \ .
\en
The analytic solutions used to compare with hydrodynamic simulations in Sections~\ref{sec:iso-wave-sim} and \ref{sec:adi-magneto-sim-EdS} were obtained by applying this reduction to
equations~\eqref{eq:drhoc-analytic-magnetosonic-only-Au} and \eqref{eq:du-analytic-magnetosonic-only-Au}.

\section{Numeric solutions}
\label{sec:numeric-scipy-solutions}

\begin{figure*}
\includegraphics[trim= 0 20 0 10]{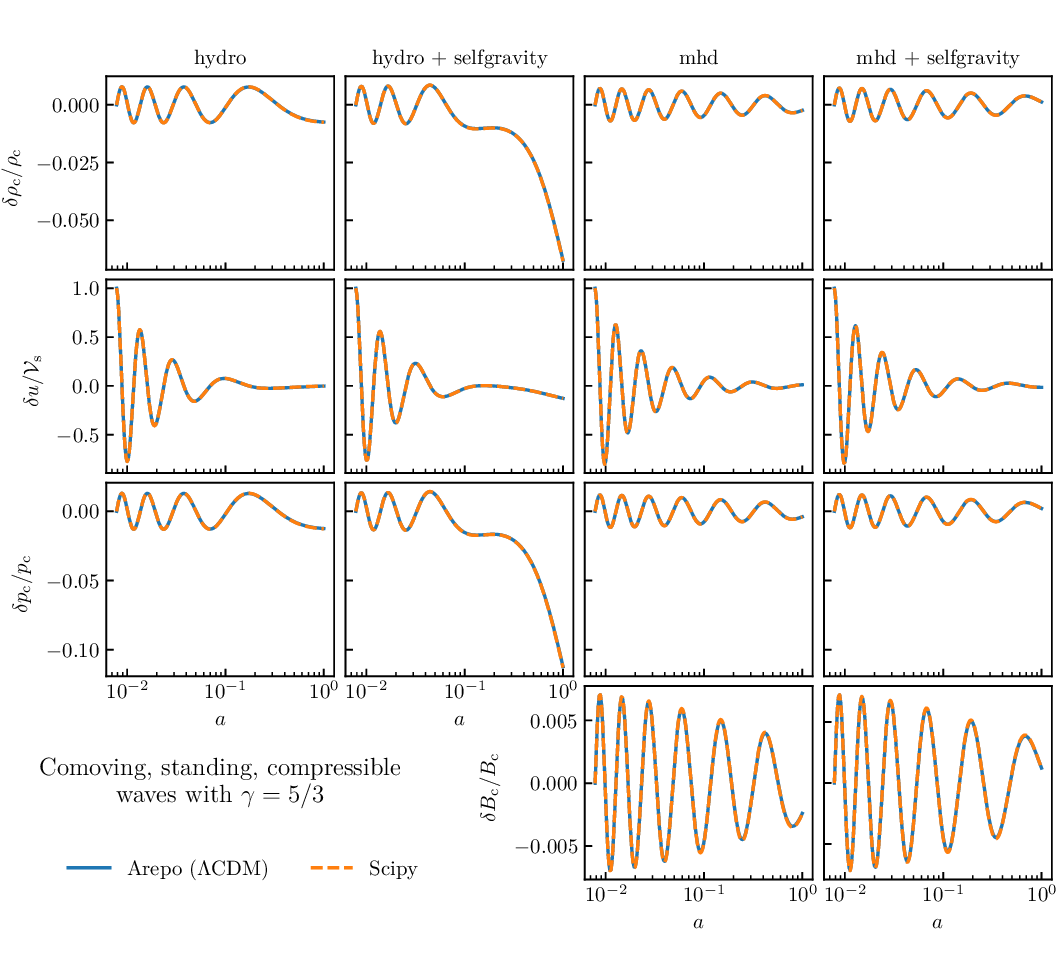}
\caption{Comparison between a numeric solution of the linearized equations (using \textsc{scipy}) and an \textsc{arepo} simulation of
a compressible, standing perturbation in a gas with adiabatic index $\gamma=5/3$ residing in a $\Lambda$CDM universe (Sections~\ref{sec:numeric-scipy-solutions}
and \ref{sec:adi-magneto-sim-LCDM}).
A comparison
with Fig.~\ref{fig:compressible-gamma-5/3-EdS} reveals that one difference between $\Lambda$CDM and EdS is that the wave performs more oscillations in $\Lambda$CDM than in EdS. The explanation for this is probably (at least partially) that the cosmic time duration of the $\Lambda$CDM simulation is roughly 40 percent longer than the EdS simulation.
}
\label{fig:compressible-gamma-5/3-LCDM}
\end{figure*}

The analytic wave solutions presented in Sections~\ref{sec:analytic-alfven-wave} and \ref{sec:analytic-magnetosonic-wave} are restricted
to cosmologies where the evolution of the scale factor takes
the simple EdS form. Here we treat the more general $\Lambda$CDM  model
by solving the first order ODE systems numerically.

In terms of the parameter $\dw = a\delta u$
the linearized Alfvén wave equations are
\be
    \pder{}{a}\f{\dBc}{\Bc} = \f{\ui k}{a^2 \dot{a}} \dw \ , \\
    \pder{\dw}{a} = \f{\ui k\vao^2}{a \dot{a}} \f{\dBc}{\Bc} \ ,
\en
while the compressible, linearized equations are
\begin{align}
    \pder{}{a}\f{\drhoc}{\rhoc} &= - \f{\ui k}{a^2\dot{a}}\dw\ ,\\
    \pder{\dw}{a} &=
    -\f{\ui k}{\dot{a}} \left(\f{\cado^2}{a^{3(\gamma-1)}} + \f{\vao^2-\vgo^2}{a}\right)
    \f{\drhoc}{\rhoc} \ .
\end{align}
In these equations $\dot{a}$ is given by the Friedmann equation (equation~\ref{eq:FLRW}).
We use Scipy's ODE solver to solve these initial value problems.

We show solutions for compressible waves with $\drhoc(\ai)=0$
with $\Omega_{\mathrm{m}}=0.3$ and $\Omega_\Lambda = 0.7$
in Fig.~\ref{fig:compressible-gamma-5/3-LCDM}. The \textsc{scipy} solution for the Alfvén wave in $\Lambda$CDM is not presented in the paper but we make our script available online.

\section{Example code testing with Arepo}
\label{sec:arepo-examples}

We use the analytic solutions derived in
Sections~\ref{sec:analytic-alfven-wave} and
\ref{sec:analytic-magnetosonic-wave} to compare with
results obtained from simulations performed with \textsc{arepo} \citep{Springel2010,Weinberger2020}. We use the MHD implementation described in \citet{Pakmor2011,Pakmor2013} and the improvements described in \citet{Pakmor2016}.

The simulations are performed on a domain of length
$L$. The waves are initialized with dimensionless amplitude $A_u = 10^{-6}$
and wavenumber $k L=2\upi$ at redshift $127$ (corresponding to $\ai=1/128$) unless otherwise specified. The standing wave simulations are initialized by perturbing the velocity only with $\du(\ai)=A_u\vao\cos(k x)$ for Alfvén waves and $\du(\ai)=A_u\cado\cos(k x)$ for compressible waves. The traveling waves
are initialized using the analytic solutions at $a=\ai$.
The amplitudes shown in the figures have been
divided by the value of $A_u$ such that the velocity amplitude appears as 1 at $a=\ai$.

Most of our simulations are performed in 1D on a static, equidistant grid.
In order to better test the Voronoi grid
in \textsc{arepo}, the convergence studies presented in Figures~\ref{fig:alfven-sim-standing}, \ref{fig:alfven-sim-traveling} and \ref{fig:compressible-gamma-4/3-traveling} are however performed in 2D.
These 2D simulations have size $L_x\times L_y$ with $N_x\times N_y$ mesh generating points where we set $N_y = 4$ and $L_y = L_x N_y/N_x$. The Voronoi generating points are given by a regular Cartesian, uniform grid where every second row has been displaced by $0.45 L_x/N_x$. This procedure creates an elongated, hexagonal grid with roughly uniform resolution along each
direction.

\subsubsection{Periodic self-gravity in 1D}

\textsc{arepo} normally employs a sophisticated, so-called TreePM gravity
solver which combines an oct-tree algorithm for short-range forces with a particle-mesh (PM) algorithm for long range forces (\citealt{Springel2005,Springel2010}, see \citealt{Weinberger2020} for a recent discussion of the \textsc{arepo} implementation). The implementation is however multi-dimensional
and cannot be used for 1D simulations. While it would be possible to perform the wave tests in 3D, we decided to instead implement a simple 1D Poisson solver. In this implementation, we discretize equation~\eqref{eq:poisson-comoving} using finite differences on an equidistant grid and obtain the potential by solving the resulting matrix
equation (using GMRES from the \textsc{GSL} library, see
\citealt{GoughGSL}). We then use centered finite differences to find the gravitational
acceleration. The implementation is limited to equidistant grids but
suffices for our purposes.

\subsubsection{Cosmological source term when $\gamma\neq1$ and $\gamma\neq5/3$}

The comoving version of \textsc{arepo} normally assumes either $\gamma=1$ or
$5/3$. The tests with $\gamma=4/3$ in Section~\ref{sec:compressible-gamma-4/3-sim} therefore required implementing an extra
source term MHD in the energy equation. We provide the details in Appendix~\ref{app:transformation-of-equations} with the extra term in the energy equation given in equation~\eqref{eq:comoving-energy-with-fewer-sources}.

\subsection{Comoving Alfvén waves}
\label{sec:alfven-sim}

\subsubsection{Standing Alfvén waves}
\label{sec:alfven-sim-standing}

The analytic solution for a linearly polarized Alfvén wave is
derived in Section~\ref{sec:analytic-alfven-wave}
with the result for the Fourier mode amplitudes given by equations~\eqref{eq:analytic-alf-dBc-osc} and \eqref{eq:analytic-alf-du-osc}.
When the initial perturbation is such that $\dBc(\ai)=0$, the solution in real space can be written
\be
    \f{\dBc(x, a)}{\Bc}
    &=&
    -A_u\left(\f{a}{\ai}\right)^{-1/4}
    \f{\sqrt{\ai}\Oma}{\kappa}
    \sin(\psi) \sin\left(k x\right) , \,\quad
    \label{eq:analytic-alf-dBc-osc-real}
    \\
    \f{\du(x, a)}{\vao} &=&
    A_u \left(\f{a}{\ai}\right)^{-3/4}
    \left(
    \cos\left(\psi\right)
    -
    \f{\sin\left(\psi\right)}{4\kappa}
    \right)\cos\left(k x\right) \ ,
    \label{eq:analytic-alf-du-osc-real}
\en
where $\Oma$, $\kappa$ and $\psi$ are defined in equations~\eqref{eq:Omega-definitions}, \eqref{eq:alfven-kappa} and \eqref{eq:psi-alfven}, respectively.

We use this solution to initialize a simulation with $\Oma=\upi$ and
show the amplitude evolution of the wave in
Fig.~\ref{fig:alfven-sim-standing-example}. The \textsc{arepo} solution is seen to closely follow the theoretical expectation.

A more detailed understanding of the frequency and amplitude error found
using \textsc{arepo} can be found by performing a series of simulations
at various spatial resolutions. We perform such a study in 2D where we expect second order convergence from the space and time discretization in \textsc{arepo} \citep{Pakmor2016}. We also use a stronger magnetic field, $\Oma=2\upi$ in order to have many oscillations occurring between $\ai=1/128$ and $a=1$.

The result of this convergence study is shown in Fig.~\ref{fig:alfven-sim-standing}. In the left panel we show $\dBc/\Bc$ multiplied with
$a^{1/4}$ and $\du/\vao$ multiplied with $a^{3/4}$ as a function of $\ln a/\ai$.
The advantage of plotting the solution this way is that it
makes the correct solution appear as simple sinusoidal functions. This makes it
easy to visually distinguish the numeric decay of the wave from the physical one. In addition to this visual assessment, we also fit the
\textsc{arepo} data with functions of the form
\be
y(x) = C \sin\left(\kappa_\mathrm{num} x + D\right)\ue^{-\Gamma_\mathrm{num} x/2}
\label{eq:fitting-formula}
\ ,
\en
where $x=\ln (a/\ai)$ and $y = \dBc a^{1/4}$ or $y=\du \,a^{3/4}$ are values
obtained from the simulations. Here $C$, $D$, $\kappa_\mathrm{num}$
and $\Gamma_\mathrm{num}$ are fitting parameters. As evident in the left panel of Fig.~\ref{fig:alfven-sim-standing} such fits provide an excellent match to the simulation
data.
The error in the frequency is shown in the top right panel
of Fig.~\ref{fig:alfven-sim-standing} (here $\kappa_\mathrm{num}$ is the
frequency found in the simulation data and $\kappa$ is the theoretical expectation). The numeric decay rate, $\Gamma_\mathrm{num}/2$, is shown in
the bottom right panel of Fig.~\ref{fig:alfven-sim-standing}.
The frequency error is seen to converge at second order while the numeric decay rate converges at third order.

The time steps in \textsc{arepo} are performed in units of $\ln a$.
The erroneous power law damping can therefore be understood as a
constant amplification factor applied at each time step equidistantly spaced
in $\ln a$ (see e.g. chapter 2 in \citealt{Durran2010} for the definition of
the amplification factor).
Alternatively, this damping can be understood as arising from a redshift
dependent numerical viscosity coefficient.
In order to make this interpretation, we amend our analytic solution for Alfvén
waves to include Navier-Stokes viscosity. The derivation is given
in Appendix~\ref{app:alfven-with-visc} with the analytic solution given in
Equations~\eqref{eq:analytic-alf-dBc-osc-visc} and
\eqref{eq:analytic-alf-du-osc-visc}. The solution shows that the numerical decay given by
$\Gamma_\mathrm{num}/2$ can be described as arising due to a numerical viscosity
coefficient $\eta_\mathrm{num} = \eta_{\mathrm{num},0}\, a^{-5/2}$ where
$\eta_{\mathrm{num},0}$ is the numerical viscosity coefficient at $z=0$.
The relation $\Gamma_\mathrm{num} = \eta_{\mathrm{num},0}\, k^2/(\rhoc H_0)$
allows us to find the numerical viscosity coefficient from the
fits shown in Fig.~\ref{fig:alfven-sim-standing}. We find
\be
    \eta_\mathrm{num} \approx 0.072 \times \f{\rhoc H_0}{k^2}
    \left(\f{N_x}{16}\right)^{-2.99} a^{-5/2} \ ,
\en
which shows how the numerical viscosity, $\eta_\mathrm{num}$, depends on scale factor and resolution.
While this numerical viscosity picture describes the amplitude damping
it does not fully explain the simulation
results. Indeed, while a physical viscosity would decrease the
wave frequency (see Appendix~\ref{app:alfven-with-visc}),
low resolution \textsc{arepo} solutions are instead seen to
display a higher frequency than the analytic solution.
This means that \textsc{arepo}'s
update procedure is accelerating rather than
decelerating (see e.g. chapter 2 in \citealt{Durran2010} for definitions
of this terminology).

\subsubsection{Traveling Alfvén waves}
\label{sec:alfven-sim-traveling}

We use the eigenmode for a traveling Alfvén wave derived in Section~\ref{sec:analytic-alfven-wave} to initialize \textsc{arepo} simulations.
Equations~\eqref{eq:alfven-dBc-eigenmode} and \eqref{eq:alfven-du-eigenmode}
are Fourier amplitudes so we need to construct the solution in real space.
A wave traveling to the right (left)
is simply found by multiplying the minus (plus) solution by
$\ue^{\ui k x}$ and taking the real part. We find it useful to write the solution using the amplitude
$A_u = C \ai^{-3/4} \ue^{\pm \ui \kappa \ln \ai}$.
The wave traveling to the right is then given by
\be
    \f{\dBc}{\Bc} &=& A_u\f{\sqrt{\ai}}{4\Oma} \left(\f{a}{\ai}\right)^{-1/4}
    \Big[\sin\left(k x- \psi\right) - 4\kappa \cos\left(k x- \psi\right)
    \Big] \ , \quad\quad
    \label{eq:alfven-dBc-traveling} \\
    \f{\du}{\vao} &=&
    A_u \left(\f{a}{\ai}\right)^{-3/4} \cos\left(k x- \psi\right) \ ,
    \label{eq:alfven-du-traveling}
\en
where $A_u$ is the dimensionless velocity amplitude. We take $\Oma=\upi$ and perform 2D simulations with $N_x=8$, 16, 32 and 64.
The simulations are initialized such that exactly $n$ periods elapse
between $\ai$ and $a=1$. This is done by setting $\ai=\ue^{-2 n \upi/\kappa}$
where $n$ is the number of times the wave traverses the box.
We choose $n=2$
such that $\ai\approx 0.01808$ for this test.

\begin{figure*}
\includegraphics[trim = 0 20 0 0]{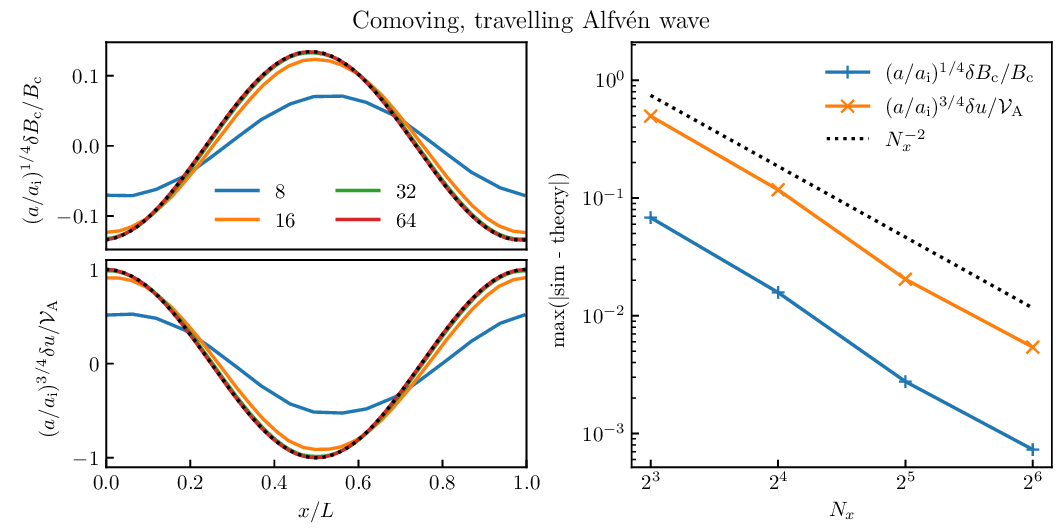}
\caption{Convergence test of a traveling Alfvén wave in EdS cosmology (Section~\ref{sec:analytic-alfven-wave} and \ref{sec:alfven-sim-traveling}). The wave is initialized at $\ai=\ue^{-2\upi n/\kappa}$ with $n=2$ such that it has traveled twice across the simulation domain by $a=1$. Left panel: the \textsc{arepo} simulation data for $\dBc$ (top) and $\du$ (bottom) is shown as a function of $x$ at $a=1$. The amplitudes are appropriately scaled with powers of $a/\ai$ such that the initial condition (shown with a dashed black line) coincides with the analytic solution at $a=1$ despite the wave decaying in amplitude while it travels. The extra decay found in the lower resolution simulations is therefore due to numerical dissipation. The analytic solution is given by equations~\eqref{eq:alfven-dBc-eigenmode}
and \eqref{eq:alfven-du-eigenmode} (Fourier amplitudes) or equations~\eqref{eq:alfven-dBc-traveling}
and \eqref{eq:alfven-du-traveling} (in real space). Right panel: The maximum of the absolute difference between \textsc{arepo} simulation and analytic solution at $a=1$ converges at second order.}
\label{fig:alfven-sim-traveling}
\end{figure*}

The traveling wave makes it easy to distinguish between frequency and amplitude errors, as evident in the left column of Fig.~\ref{fig:alfven-sim-traveling} where we show the wave profiles at $a=1$. For instance, we observe that the waves are traveling too fast in the low resolution \textsc{arepo} simulations (i.e. $\kappa_{\mathrm{num}}>\kappa$).
The (absolute, maximum) difference between theory and simulation converges at second order, as shown on the right in Fig.~\ref{fig:alfven-sim-traveling}.

\subsubsection{Alfvén waves with a deliberately introduced bug in \textsc{arepo}}
\label{sec:alfven-wave-with-intentional-bug}

Many types of programming errors in the $a$-factors will lead to nonsensical
results or perhaps even premature termination of a simulation (if infinities
or nans develop). Here we illustrate that some errors in $a$-factors can lead to subtle effects that are not so easily found by the code developer.
This exercise is done to show that sensible looking simulations that do not crash are not necessarily bug-free.

We use for our example the MHD flux calculation as developed by
\citet{Pakmor2013} and described
in their section 2.2. In their algorithm the magnetic field is scaled as
$\vec{B}'_\mathrm{c}=\vec{B}_\mathrm{c}/\sqrt{a}$ (their step \emph{ii}),
a Riemann solver is used to calculate the flux, $\vec{F}'_B$ (their step
\emph{iii}), and finally the scaling of the magnetic field is reverted by
transforming the flux: $\vec{F}_B = \vec{F}'_B \times \sqrt{a}$
(their step \emph{iv}).

We stress that this implementation works in \textsc{arepo}. Let us, however,
imagine that something went wrong in step \emph{iv} such that either
$\vec{F}_B = \vec{F}'_B$ (the flux was never rescaled, we will call this
hypothetical bug "Bug 1") or $\vec{F}_B = \vec{F}'_B \times a$ (the flux was
rescaled by the wrong $a$-factor, we will call this hypothetical bug "Bug 2").

\begin{figure}
\includegraphics[trim = 0 20 0 0]{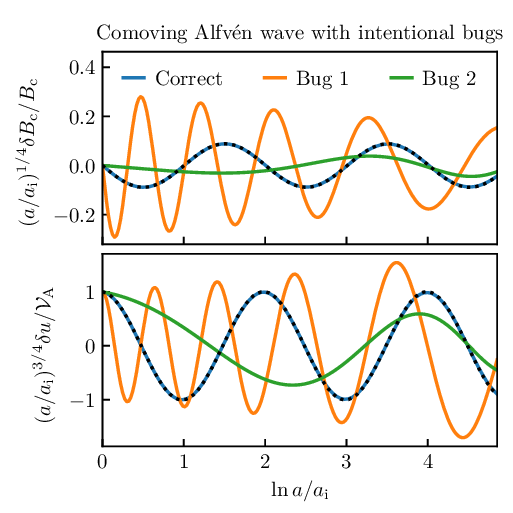}
\caption{Alfvén wave in EdS cosmology with \emph{intentionally} introduced bugs in the MHD flux calculation (Section~\ref{sec:analytic-alfven-wave} and \ref{sec:alfven-wave-with-intentional-bug}). Bug 1 (2) has an extra factor of $a^{-1/2}$ ($a^{1/2}$) in the flux. This leads to a too high (Bug 1) or too low (Bug 2) frequency as compared
to the correct result. The analytic theory is shown with a dashed black line. The example is meant to illustrate how an error in the MHD flux calculation can be detected.}
\label{fig:alfven-sim-with-bug}
\end{figure}

We \emph{intentionally} introduce these two types of bugs in the \textsc{arepo}
source code and perform standing, Alfvén wave simulations using the same
parameters and field strength as in Fig.~\ref{fig:alfven-sim-standing-example},
i.e., $\Oma=\upi$.
The two types of bugs lead to the wrong frequency and
amplitude evolution of the Alfvén wave. This is  evident in
Fig.~\ref{fig:alfven-sim-with-bug} where we compare with the original (and
correct) implementation. While problems are immediately discovered using the Alfvén wave test, it is less obvious how these bugs would manifest themselves in a so-called "full physics" cosmological
simulation. One merit of the Alfvén wave test is thus that it
catches even subtle errors.

\begin{figure*}
\includegraphics[trim = 0 20 0 0]{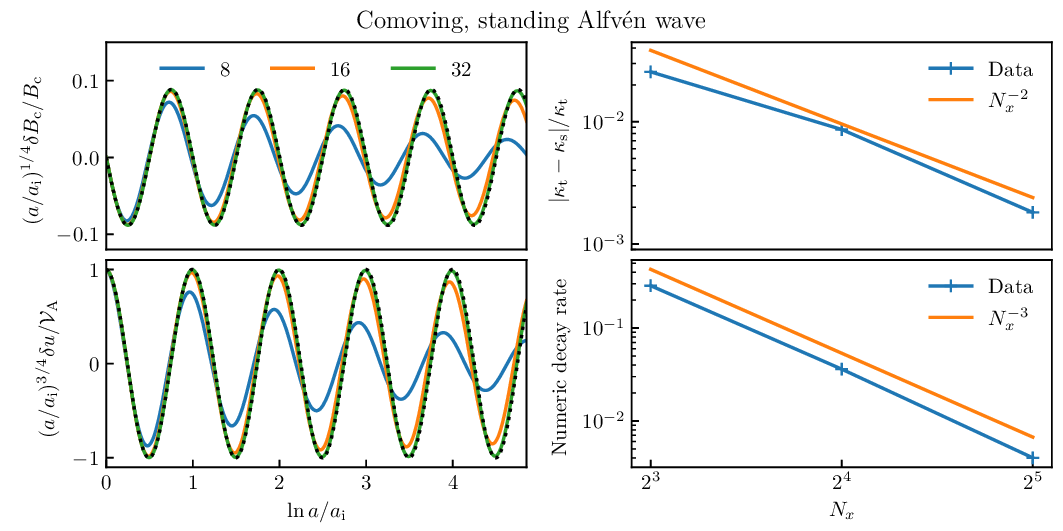}
\caption{Convergence test of a standing Alfvén wave in EdS cosmology (Section~\ref{sec:analytic-alfven-wave}
and \ref{sec:alfven-sim-standing}).
Left panel: the amplitudes of $\dBc$ (top) and $\du$ (bottom) scaled with appropriate powers of $a$ such that the solutions appear sinusoidal when plotted versus $\ln a/\ai$. The decay in amplitude is therefore due to numerical dissipation only. The analytic solution is shown with solid black lines while fits to equation~\eqref{eq:fitting-formula}
are shown with dotted black lines. The labels indicate the number of cells in the $x$-direction. With $N_x=32$ the
\textsc{arepo} simulation data coincides visually with the analytic solution while $N_x=16$ ($N_x=8$) has (significant) amplitude and frequency error.
Right panel: Convergence of the frequency (top) and numerical decay rate (bottom). }
\label{fig:alfven-sim-standing}
\end{figure*}

\subsection{Comoving compressible waves}
\label{sec:compressible-simulations}

\subsubsection{Waves with $\gamma=4/3$}
\label{sec:compressible-gamma-4/3-sim}

The baryonic gas in cosmological simulations is normally assumed to obey
an adiabatic equation of state with $\gamma=5/3$ (e.g. \citealt{Weinberger2020}). The value chosen in
the following test, $\gamma=4/3$, has the curious feature that the thermal, magnetic and gravitational forces decay at the same rate. This value of $\gamma$ describes a gas with a relativistic equation of state. While a relativistic component is expected in the form of cosmic rays accelerated at e.g. structure formation shocks (e.g. \citealt{Pfrommer2006}), the equations of ideal MHD that we use here do not correctly include all the relevant terms
(see e.g. \citealt{Pfrommer2017} for the equations for fluid cosmic rays coupled to thermal gas in a comoving frame). Nevertheless, we argue that the following examples
are useful because their analytic solutions are given in terms of simple, elementary functions. This is an advantage compared to the $\gamma=1$ or $\gamma=5/3$ solutions
presented in Section~\ref{sec:comp-gamma-not-4/3} whose evaluation require
Bessel functions of imaginary order. In addition, it is possible to set up both standing and traveling compressible waves for $\gamma=4/3$. This is only
possible because the wave eigenmode is known (see Section~\ref{sec:comp-gamma-4/3}).

\subsubsection*{Standing waves}

\label{sec:compressible-gamma-4/3-sim-standing}

We take $\Oms = \upi$ with $\Oma=\upi$ for MHD simulations and
$\Omg=\upi/2$ for simulations with self-gravity (with $\Oma=0$ when MHD is turned off and $\Omg=0$ when self-gravity is turned off).
The general analytic solution for the Fourier amplitudes for this problem are
given by equations~\eqref{eq:analytic-drho-standing-gamma-4/3} and
\eqref{eq:analytic-du-standing-gamma-4/3}
in Section~\ref{sec:comp-gamma-4/3}. When $\drhoc(\ai)=0$, the analytic
solution in real space can be written
\be
    \f{\drhoc}{\rhoc} =
    A_u \left(\f{a}{\ai}\right)^{-1/4}
    \f{\Oms \sqrt{\ai}}{\kappa}
    \sin(\psi) \sin(k x)\ , \quad\quad
    \label{eq:analytic-drho-standing-gamma-4/3-real}
    \\
    \f{\du}{\cado} =
    A_u \left(\f{a}{\ai}\right)^{-3/4}
    \left(
    \cos\left(\psi\right)
    -
    \f{\sin\left(\psi\right)}{4\kappa}
    \right)\cos(k x) \ ,
    \label{eq:analytic-du-standing-gamma-4/3-real}
\en
where $\kappa$ is defined in equation~\eqref{eq:kappa-gamma-4/3} and
$\psi = \kappa \ln \left({a}/{\ai}\right)$.
The comparison
between theory and simulation is presented in
Fig.~\ref{fig:compressible-gamma-4/3-standing}
where we find excellent agreement. We note that the solutions have been scaled and plotted versus $\ln a/\ai$ such that they appear as sinusoidal functions with constant amplitude
(in the same way as we did for the Alfvén wave in Fig.~\ref{fig:alfven-sim-standing}).

Consistent with equation~\eqref{eq:kappa-gamma-4/3}, the magnetic field is seen to increase the frequency of the wave while self-gravity decreases the frequency.
The same qualitative behaviour is seen in standard, non-comoving ideal MHD (e.g. \citealt{Pringle2014}). The physical mechanism is that the magnetic pressure acts as a restoring force in unison with the thermal pressure
while self-gravity acts in the opposite direction and therefore decreases the frequency. In addition to the changes in frequency, self-gravity also slightly increases the amplitude of the density perturbation
relative to simulations without self-gravity (this is clearly visible in the
hydrodynamic simulations) while the magnetic field slightly decreases the amplitude.

\subsubsection*{Traveling waves}

\label{sec:compressible-gamma-4/3-sim-traveling}

Here we initialize an \textsc{arepo} simulation with
waves traveling to the right. The analytic solution was derived in
Section~\ref{sec:comp-gamma-4/3} and can be written in real space as
\be
    \f{\drhoc}{\rhoc} &=& A_u\f{\Oms\sqrt{\ai}}{4\sigma^2} \left(\f{a}{\ai}\right)^{-1/4}
    \left[4\kappa \cos\left(k x- \psi\right) -
    \sin\left(k x- \psi\right)
    \right] \ , \quad \ \,
    \label{eq:magneto-drhoc-traveling} \\
    \f{\du}{\cado} &=&
    A_u \left(\f{a}{\ai}\right)^{-3/4} \cos\left(k x- \psi\right) \ ,
    \label{eq:magneto-du-traveling}
\en
where $\sigma$ was defined in equation~\eqref{eq:sigma-definition}.
We set $\Oms=\Oma=\upi$ and $\Omg=0$ and start the simulation at $\ai=\ue^{-2 n \upi/\kappa}$ where $\kappa$ is defined in equation~\eqref{eq:kappa-gamma-4/3}
and $n$ is the number of times the wave travels through the box. We set $n=3$
such that $\ai\approx 0.01427$ for this test.

\begin{figure*}
\includegraphics[trim = 0 20 0 0]{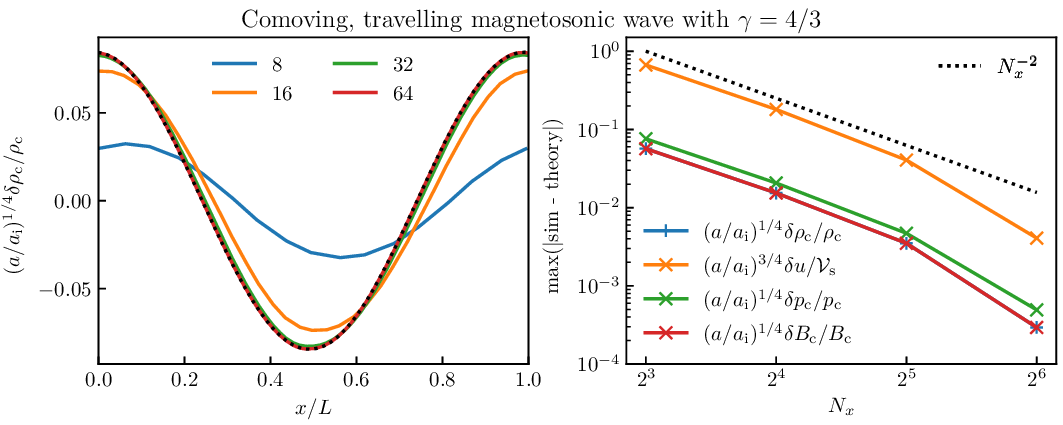}
\caption{Same as Fig.~\ref{fig:alfven-sim-traveling} but here for
a traveling magnetosonic wave with $\gamma=4/3$ and $\Oms=\Oma=\upi$
which has traveled three times through the box (Sections~\ref{sec:comp-gamma-4/3} and \ref{sec:compressible-gamma-4/3-sim-traveling}). The analytic solutions are given by equations~\eqref{eq:magneto-drhoc-fourier-traveling-gamma-4/3} and \eqref{eq:magneto-du-fourier-traveling-gamma-4/3} (Fourier amplitudes) and equations~\eqref{eq:magneto-drhoc-traveling} and \eqref{eq:magneto-du-traveling} (in real space).}
\label{fig:compressible-gamma-4/3-traveling}
\end{figure*}

We perform 2D simulations and vary $N_x$ in powers of 2 between 8 and 64. We show the results in
Fig.~\ref{fig:compressible-gamma-4/3-traveling} with the wave profiles
on the left and the convergence results on the right. As for the traveling
Alfvén wave, we find second order convergence of the (absolute, maximum) difference between theory and simulation.

\subsubsection{Isothermal waves}
\label{sec:iso-wave-sim}

We next consider waves with an isothermal equation of state (i.e. $\gamma=1$).
The interesting feature for an isothermal equation of state is that the overall
strength of the thermal restoring force of the wave does not decay as a
function of redshift. This means that it is possible to set up a simulation
which is initially dominated by magnetic and/or gravitational forces and then
transitions to be dominated by thermal forces. By carefully choosing the
parameters ($\Oms$, $\Oma$ and $\Omg$), it is possible to have the time period
where gravity dominates long enough that a strong modification is seen (relative
to simulations without self-gravity) but short enough that the wave remains
firmly in the linear regime. We have found the parameters $\Oms=2\upi$ with
$\Oma=\upi$ when MHD is turned on and $\Omg=\pi/2$ when self-gravity is turned on to result in this behaviour.

We show the results of simulations with these parameters in Fig.~\ref{fig:compressible-gamma-1-EdS}. The hydrodynamic solution with self-gravity is shown in the second column of Fig.~\ref{fig:compressible-gamma-1-EdS}. This solution has a decreased frequency and a very large density amplitude compared to the hydrodynamic solution without self-gravity (its larger by a factor of almost five). The extra growth in density amplitude is seen to occur at
early times. This is in good agreement with equation~\eqref{eq:Jeans-criterion}
which predicts that gravity dominates for $a<1/16$. The magnetic field in the MHD plus self-gravity simulation dominates gravity  at all times (since $\Oma>\Omg$ and the forces decay at the same rate). There is therefore only a slight
increase in the density amplitude relative to the MHD simulation without
self-gravity (compare the third and fourth columns in Fig.~\ref{fig:compressible-gamma-1-EdS}).

The relevant theory for these simulations was derived in Section~\ref{sec:analytic-magnetosonic-wave}
with the Fourier amplitudes given by equations~\eqref{eq:drhoc-analytic-magnetosonic-only-Au}
and \eqref{eq:du-analytic-magnetosonic-only-Au} in Section~\ref{sec:comp-gamma-not-4/3}.
The solution can be written in a much simpler form for the hydrodynamic problem without self-gravity. In this case, we find for $\gamma=1$ that the analytic solution for a standing wave in real space is\footnote{Unlike Alfvén waves
and $\gamma=4/3$ waves, the isothermal soundwaves are not dispersive. That is
\be
    \dot{\phi} = k \cado a^{-1} = k' \cado \ ,
\en
such that the sound wave frequency is as expected from the change in wavenumber.
}
\be
    \f{\drhoc(x,\,a)}{\rhoc} &=& A_u \sqrt{\f{\ai}{a}} \sin\left(\phi\right)\sin\left(k x\right)\ ,
    \label{eq:hydro-drhoc-gamma1}\\
    \f{\du(x,\,a)}{\cado} &=& A_u \left[\sqrt{\f{\ai}{a}} \cos\left(\phi\right) -
    \f{\sqrt{\ai}}{2\Oms a} \sin\left(\phi\right)\right] \cos\left(k x\right)\ ,\quad
    \label{eq:hydro-du-gamma1}
\en
where
$
\phi = 2\Oms \left(\sqrt{a} - \sqrt{\ai}\right) 
$.
As evident in Fig.~\ref{fig:compressible-gamma-1-EdS}, the simulations agree
very well with the evolutions predicted by the analytic theory.

\subsubsection{Adiabatic waves in EdS}
\label{sec:adi-magneto-sim-EdS}

We proceed by studying adiabatic waves by setting $\gamma=5/3$. This is
the value used in almost all cosmological MHD simulations and is consequently
a very relevant test case. In addition, $\gamma=5/3$ has the interesting feature that the thermal force decays
\emph{faster} than the gravitational and magnetic forces. This means that
it is possible to set up a wave which is initially gravitationally stable but
transitions to gravitational instability once the thermal pressure drops below
a critical value. We use equation~\eqref{eq:Jeans-criterion} to set up the parameters in precisely this way, i.e., the hydrodynamic simulation with
self-gravity goes unstable near the end of simulation. We choose the value of $\Oma$ such that the MHD simulation with self-gravity is stable at all times. Choosing the parameters this way gives a qualitative difference between the hydrodynamic and MHD tests including self-gravity (rather than just a quantitative change in frequency and amplitude). Given this motivation, we take $\Oms = \upi/5$ with $\Oma=\upi$ for MHD simulations and $\Omg=\upi/2$ for simulations with self-gravity.

A comparison between \textsc{arepo} simulations and analytic theory is shown in
Fig.~\ref{fig:compressible-gamma-5/3-EdS}. We note the excellent agreement between the two and the onset of gravitational instability in the hydrodynamic simulation with self-gravity (second column in Fig.~\ref{fig:compressible-gamma-5/3-EdS}) as well as the absence of gravitational instability in the MHD plus self-gravity simulation (fourth column).

The analytic solutions used for the comparison are given by equations~\eqref{eq:drho-besselY-solution} and \eqref{eq:du-besselY-solution} in Section~\ref{sec:comp-gamma-not-4/3}. The hydrodynamic wave without magnetic fields
and self-gravity can be written in a much simpler form. For $\gamma=5/3$, the solution with $\drhoc(\ai)=0$ becomes\footnote{The adiabatic soundwaves are dispersionless. We expect $\omega_{\mathrm{s}} = k'\cad = k/a \cado a^{-3(\gamma-1)/2}$
which for $\gamma=5/3$ becomes $\omega_{\mathrm{s}} = k \cado a^{-2}$.
Calculating the time derivative of $\varphi$ we find
\be
    \dot{\varphi} = k \cado a^{-2} \ ,
\en
in agreement with this prediction.
}
\be
    \f{\drhoc(x,\,a)}{\rhoc} &=& A_u \ai \sin\left(\varphi\right)\sin\left(k x\right)\ ,
    \label{eq:drhoc-gamma5/3}
    \\
    \f{\du(x,\,a)}{\cado} &=& A_u \f{\ai}{a} \cos\left(\varphi\right)
    \cos\left(k x\right)\ ,\quad
    \label{eq:du-gamma5/3}
\en
where
\be
    \varphi = 2\Oms \left(\f{1}{\sqrt{\ai}} - \f{1}{\sqrt{a}}\right) \ .
\en

\subsubsection{Adiabatic waves in $\Lambda\mathrm{CDM}$}
\label{sec:adi-magneto-sim-LCDM}

We perform \textsc{arepo} simulations of adiabatic waves in $\Lambda$CDM cosmology.
We set $\Omega_{\mathrm{m},0}=0.3$ and $\Omega_{\Lambda, 0} = 0.7$ in equation~\eqref{eq:FLRW-simple} and keep all other parameters as
in the preceding section where we used EdS cosmology. It is important to note that we do not include any dark matter particles in the simulation. That is, the test only includes dark matter as a uniform background which influences the evolution of $a$ via the Friedmann equation (equation~\ref{eq:FLRW-simple}).

We do not have an analytic solution for $\Lambda$CDM and instead compare the \textsc{arepo} simulations with the numeric solution
described in section~\ref{sec:numeric-scipy-solutions}. This comparison between \textsc{arepo} simulation and \textsc{scipy} solution is shown in Fig.~\ref{fig:compressible-gamma-5/3-LCDM}.
It it interesting to compare the $\Lambda$CDM results (Fig.~\ref{fig:compressible-gamma-5/3-LCDM}) with those found for EdS (Fig.~\ref{fig:compressible-gamma-5/3-EdS}). The most notable difference between waves in the two cosmologies is that the waves oscillate
more times in $\Lambda$CDM than in EdS. The main reason for this is likely that
the cosmic time duration between $\ai=1/128$ and $\ai=1$ is roughly 40 percent
longer in $\Lambda$CDM than it is in EdS.

\section{Discussion}
\label{sec:discussion}

With the rapid increase in the complexity of codes for computational astrophysics, the code development is increasingly becoming
a continued and collaborative effort. In order to avoid introduction of errors (so-called bugs) and to ensure that
a given code version is fit for production runs, developers are using version control systems (primarily \textsc{git} version control, see e.g. \citealt{chacon2014pro}) and adapting software engineering practices such as automated testing. This practice allows for continuous integration of new features into the main code version (with \textsc{git} by making a pull-request from a development branch into the main branch).

However, an essential ingredient necessary for building confidence in this process is that the test suite covers all the intended use cases.
Since comoving MHD codes are generally tested by using test problems developed
for standard MHD codes (by setting $a=1$ and $\dot{a}=0$),
we argue in this paper that this criterion for a successful testing framework is often not fulfilled. This does not necessarily mean that published results relying on cosmological MHD simulations are wrong\footnote{As evident from the fact that \textsc{arepo} passed all tests in this paper.} but rather that the underlying code base could be vulnerable to the introduction of errors in the source code (a regression).

The purpose of the present paper is to remedy this issue by introducing
new tests specifically designed to help find programming errors related to $a$ and $\dot{a}$. These new tests consist of analytic solutions for
Alfvén and magnetosonic waves in EdS cosmology (derived in Section~\ref{sec:analytic-solutions}) and numeric solutions for those waves in $\Lambda$CDM (described in Section~\ref{sec:numeric-scipy-solutions}). We have used these hydromagnetic solutions to compare with the MHD code \textsc{arepo} in Section~\ref{sec:arepo-examples}. Reassuringly, we have in all cases found excellent agreement with \textsc{arepo}.

Many of the mistakes possible when introducing scale factors in a code will
lead to obviously meaningless results or premature termination of the code.
However, it is also possible to have implementation mistakes that cause more
subtle errors. We make this point explicit by intentionally introducing bugs
in \textsc{arepo} in Section~\ref{sec:alfven-wave-with-intentional-bug}. As
seen in Fig.~\ref{fig:alfven-sim-with-bug}, the intentionally wrong
simulations are able to proceed until $a=1$ without the introduction of
obviously wrong values in the solution. However, comparison with the analytic
solution immediately reveals that there is a problem.

Subtle errors are also more likely to be introduced when additional physical
effects are included in a code base. As an example, assume that the viscosity
term in comoving coordinates scales as $a^{n}$. If the programmer by mistake
instead implements viscosity with a term proportional to $a^{n+1}$ then this will
not lead to a crash. Instead, the included effective viscosity will simply be
less than the physically expected value. This type of error is difficult to find without proper tests (Berlok et al, in prep.).

In addition to error finding, it is also possible that our tests can help
distinguish the advantages and disadvantages of various methods for
numerically solving the comoving MHD equations (i.e. by studying wave
amplitude and frequency errors at a given numerical resolution as in
Figs.~\ref{fig:alfven-sim-standing}, \ref{fig:alfven-sim-traveling} and
\ref{fig:compressible-gamma-4/3-traveling}).

For the reasons outlined above, we are of the view that these tests,
specifically designed for comoving MHD, will be an useful addition to
automated testing frameworks of comoving MHD codes.

As a final remark, we note that other fluid dynamics tests could be fruitfully extended to the comoving coordinate system as well. While analytic solutions might not always be possible, high fidelity numeric solutions will be. One particular interesting but simple type of system that could be investigated is the shock tube problem in comoving coordinates.

\section*{Acknowledgements}
TB thanks the reviewer, Federico Stasyszyn, for very useful comments
which helped improve the manuscript.
TB thanks Rüdiger Pakmor and Christoph Pfrommer for sharing their expertise on cosmological simulations and Martin Sparre for the suggestion to illustrate the merits of testing by artificially introducing a code bug in \textsc{arepo}.

\subsection*{Software}
TB is grateful to Volker Springel, Rüdiger Pakmor and Rainer Weinberger for making \textsc{arepo} available.
We use the Python library \textsc{mpmath} for numeric
evaluation of Bessel functions \citep{mpmath}.

\section*{Data availability}

Python code for evaluating the analytic solutions
presented in Section~\ref{sec:analytic-solutions} and for re-computing the numeric solutions
presented in Section~\ref{sec:numeric-scipy-solutions} can be downloaded
at \url{https://github.com/tberlok/comoving_mhd_waves}.
A public version of the \textsc{arepo} code has been made
available by the main developers \citep{Weinberger2020} and
can be downloaded at \url{https://gitlab.mpcdf.mpg.de/vrs/arepo}.
We will share our setup for initializing and analyzing
\textsc{arepo} simulations upon request.

\bibliographystyle{mnras}
\bibliography{references}

\appendix

\section{Comoving  MHD tests currently in use in
the literature}

\label{sec:current-comoving tests}

We provide a short description of the tests normally used to test implementations of comoving hydrodynamics and/or MHD.

A common test case for hydrodynamics in comoving coordinates is the so-called
Zel'dovich pancake problem which is based on the analytic results by \citet{Zel'Dovich1970}. This test combines self-gravity and a hydrodynamic fluid
and follows gravitational collapse in an EdS cosmology. The Zel'dovich pancake problem has been used in many papers including \citet{Cen1992,Ryu1993,Bryan1995,Trac2004,Bryan2014} and \citet{Springel2010}. It has been also been extended to MHD in \citet{Collins2010} and \citet{Hopkins2016}. The magnetic field evolution in these simulations is compared with reference simulations which have been performed with other codes or the same code at higher resolution.

Another common test is to consider a uniform universe in which there are no
peculiar motions (see e.g. \citealt{Collins2010} and
\citealt{Ruszkowski2011}). This
reduces the comoving MHD equations to a system of ordinary differential equations which can be analytically integrated. This type of test is useful for verifying the implementation of comoving source terms but it has the limitation that it does not verify the calculation of gradients and fluxes.

The Santa-Barbara cluster consists of a set of fixed initial conditions for a
$10^{15}M_\odot$ galaxy cluster \citep{Frenk1999}. The
diagnostics for this test are global properties, projection images and
radial profiles of various gas quantities
(e.g. gas density, pressure and temperature and dark matter density).
The Santa-Barbara cluster is an already non-trivial problem and obtaining the physically correct radial profiles can be difficult (see \citealt{Springel2010} for a discussion).
The Santa-Barbara cluster was introduced in 1999 at a time when most cosmological simulations did not include magnetic fields. In recent years
it has been extended to MHD in \citet{Miniati2011,Hopkins2016} and \citet{Hopkins2016b} but a code-comparison project including MHD has not yet been done. Simulations performed by \citet{Vazza2018} have found
that very high spatial resolution is required to obtain a converged magnetic field in non-radiative MHD simulations of a galaxy clusters. We argue that this makes the Santa-Barbara cluster unfeasible for frequent regression testing of comoving MHD.

Finally, most code papers contain example applications of their codes, e.g.,
simulations of individual galaxies or galaxy clusters. These tests are useful
for stress-testing, as codes will sometimes crash in realistic applications even though they perform well on all the idealized test problems. The disadvantage of using tests that are so close to the target science is that there is rarely a universally agreed-upon correct solution
(an example is the MHD version of the Santa-Barbara cluster discussed above).

We have here only described the tests commonly used for comoving MHD. It it
therefore worth mentioning that cosmological codes typically also test their gravity implementations against an extensive list of standard tests (see e.g. \citealt{Springel2021,Dakin2021,Angulo2022}).

\section{Transformation of MHD equations to comoving coordinates}
\label{app:transformation-of-equations}

\subsection{Friedmann equations}
\label{app:friedmann}

The evolution of the scale factor, $a$, in a flat universe
depends on the cosmological parameters for the total matter density (baryonic plus dark matter), $\Omega_{\mathrm{m},0}$, radiation, $\Omega_{\mathrm{r},0}$, and dark energy, $\Omega_{\Lambda,0}$.
This dependence is given by the Friedmann equation (e.g. \citealt{Cimatti2019,Angulo2022})
\be
    \left(\f{\dot{a}}{a}\right)^2 = \f{8\upi G}{3}
    \bar{\rho}_\mathrm{tot} \ ,
    \label{eq:FLRW}
\en
where
\be
    \bar{\rho}_\mathrm{tot} =
    \left(
    \f{\Omega_{\mathrm{r},0}}{a^4} +
    \f{\Omega_{\mathrm{m},0}}{a^3}
    +
    \Omega_{\Lambda,0}
    \right) \rho_{\mathrm{crit}, 0} \ ,
    \label{eq:rho_tot}
\en
is the mean of the total density, $\rho_{\mathrm{crit}, 0} = 3H_0^2/(8\upi G)$ is the critical density, and $H_0$ is the Hubble parameter at $z=0$.
Equations~\eqref{eq:FLRW} and \eqref{eq:rho_tot} can be combined into
equation~\eqref{eq:FLRW-simple} used in the main body of the paper.

\subsection{The coordinate transformation}
We briefly review how to transform from the fixed $\vec{r}$ coordinates to the comoving $\vec{x}$ coordinates.
We consider a general function of $\vec{r}$ and $t$, say $f(\vec{r},\,t)$. From the point of point of view of the $\vec{r}$ coordinate system, the convective time-derivative is
\be
    \der{f}{t} =
    \pdertr{f} +
    \left(\dot{\vec{r}} \bcdot \delr\right) f \ ,
    \label{eq:d/dt-lab}
\en
where the partial time-derivative is at fixed position $\vec{r}$.
Given the relation, $\vec{r} = a \vec{x}$, $\dot{\vec{r}}$ can be written out as
\be
    \dot{\vec{r}} = \vec{\varv} = \dot{a}{\vec{x}} + a \dot{
    \vec{x}} \
    = \dot{a}{\vec{x}} + \vec{u} \ ,
    \label{eq:v-u-relation}
\en
where we have defined the peculiar, physical velocity, $\vec{u} \equiv a
\dot{\vec{x}}$. The convective derivative can thus also be written
\be
    \der{f}{t} =
    \pdertr{f} +
    \dot{a}\left({\vec{x}} \bcdot \delr\right) f +
    \left(\vec{u} \bcdot \delr\right) f \ .
    \label{eq:d/dt-lab-rewritten}
\en
From the point of view of the comoving coordinate system, the convective derivative is
\be
    \der{f}{t} = \pdertx{f} +
    \left(\dot{\vec{x}} \bcdot \delx\right) f
    =
    \pdertx{f} +
    \f{1}{a}\left(\vec{u} \bcdot \delx\right) f
    \ ,
    \label{eq:d/dt-comoving}
\en
where the partial time-derivative is at fixed $\vec{x}$.

We now use that the convective derivative has a unique value which does not depend on the coordinate system.
Noting that the gradient operators are simply related
\be
    \delr f =
    \f{1}{a} \delx f \ ,
    \label{eq:spaceder-relation}
\en
we can therefore equate equations~\eqref{eq:d/dt-lab-rewritten} and
\eqref{eq:d/dt-comoving} to find a relation between the partial
time-derivatives in the two coordinate systems. This yields
\be
    \pdertr{f} = \pdertx{f}
    - \f{\dot{a}}{a} \left({\vec{x}} \bcdot \delx\right) f \ .
    \label{eq:tder-relation}
\en

\subsection{Continuity equation}
With $\rho = \rhoc a^{-3}$ and $\vec{\varv}=
\dot{a}{\vec{x}} + \vec{u}$ equation~\eqref{eq:rho} becomes
\be
    \der{\ln \rhoc}{t} -
    3\der{\ln a}{t}=
    -\f{\dot{a}}{a} \delx \bcdot \vec{x} -
    \f{1}{a}\delx \bcdot \vec{u} \ .
\en
Since $\delx \bcdot \vec{x} = 3$, the terms involving $\dot{a}$
cancel out and we obtain equation~\eqref{eq:rho-comoving}.

\subsection{Momentum equation}

\label{app:momentum-equation-derivation}

The momentum equation, equation~\eqref{eq:mom}, is transformed by
substituting $p = a^{-3} p_\mathrm{c}$, $\rho = a^{-3} \rhoc$,
$\vec{B}=a^{-2} \vec{B}_\mathrm{c}$, replacing $\delr$ with $\delx/a$ and
finally multiplying through by $a^4$. This yields
\begin{multline}
    \rhoc a \left[\der{(\dot{a}\vec{x})}{t} + \der{\vec{u}}{t}
        \right]
=
    - \delx p_\mathrm{c}
    - \f{1}{a}\delx \bcdot \left(\f{\Bc^2}{2}\mathbf{1} -
    \vec{B}_\mathrm{c}\vec{B}_\mathrm{c} \right)
    - \rhoc \delx \Phi \ ,
\end{multline}
where
\be
    \der{(\dot{a}\vec{x})}{t}
    =
    \pdertx{(\dot{a}\vec{x})}
    +
    \f{1}{a}\left(\vec{u}\bcdot \delx \right)\left(\dot{a}\vec{x}\right)
    =
    \ddot{a}\vec{x} + \f{\dot{a}}{a} \vec{u} \ .
\en
We conclude that the momentum equation can be written
\begin{multline}
    \rhoc \der{\vec{u}}{t}
=
    - \f{1}{a}\delx p_{\mathrm{tot, c}}
    - \rhoc \f{\dot{a}}{a} \vec{u} - \rhoc \ddot{a}\vec{x}
    + \f{1}{a^2}\delx \bcdot \left(
    \vec{B}_\mathrm{c}\vec{B}_\mathrm{c} \right)
    - \rhoc \f{1}{a} \delx \Phi \ ,
    \label{eq:mom-comoving-tmp}
\end{multline}
where we have introduced the total pressure $p_{\mathrm{tot, c}} = \pc + \Bc^2/(2a)$.

The gravitational potential, $\Phi$, fulfills the Poisson equation
\be
    \delr^2 \Phi = 4 \upi G \rho_\mathrm{tot} \ ,
\en
where $\rho_\mathrm{tot}$ is the total density which enters the first Friedmann equation (equation~\ref{eq:FLRW}, \citealt{Angulo2022}). Thus depending on the cosmological model employed, $\rho_\mathrm{tot}$ can include contributions from both baryonic and dark matter as well as radiation and the cosmological constant.

It is conventional to split $\Phi$ into two
contributions, $\Phi = \Phi_0 + \delta \Phi$, by splitting the density
into two contributions, $\rho_\mathrm{tot} =
\bar{\rho}_\mathrm{tot} + \delta \rho_\mathrm{tot}$ where
$\bar{\rho}_\mathrm{tot}$ is the mean density
and $\delta \rho_\mathrm{tot}$ is the local deviation, i.e.,
$\delta \rho_\mathrm{tot} = \rho_\mathrm{tot} - \bar{\rho}_\mathrm{tot}$.
This can be done without loss of generality and
$|\delta \rho_\mathrm{tot}/\bar{\rho}_\mathrm{tot}|$ does not have to be
small.
The splitting yields two separate Poisson equations, i.e.,
\be
    \delr^2 \Phi_0 = 4 \upi G \bar{\rho}_\mathrm{tot} \ ,
    \label{eq:poisson-uniform}
\en
and
\be
    \delr^2 \delta \Phi = 4 \upi G \, \delta \rho_\mathrm{tot}
    = 4 \upi G \left(\rho_\mathrm{tot} - \bar{\rho}_\mathrm{tot}\right)\ .
    \label{eq:poisson-deviation}
\en
Equation~\ref{eq:poisson-uniform} can be analytically integrated because
$\bar{\rho}_\mathrm{tot}$ is constant in space. The solution is $\Phi_0 = 2\upi G \bar{\rho}_\mathrm{tot} r^2/3$. Since $\bar{\rho}_\mathrm{tot}$ is the density in a homogeneous universe,
it is given in terms of the second Friedmann equation (see e.g. \citealt{Carroll2004})
\be
    \f{\ddot{a}}{a} = - \f{4\upi G}{3} \bar{\rho}_\mathrm{tot} \ .
\en
Using this expression for the total density it is concluded that
\be
    \Phi_0 = -\f{1}{2}\f{\ddot{a}}{a} r^2
    = -\f{1}{2} a\ddot{a} x^2
\en
where $r= a x$ was used in the second step. This agrees with
equation 7.7 in \citet{Peebles1980}. The potential has the property
that
\be
    \ddot{a} \vec{x} = -\f{1}{a}\delx \Phi_0 \ ,
\en
which can be seen to eliminate
the $\rhoc \ddot{a} \vec{x}$ source term in the momentum equation (\ref{eq:mom-comoving-tmp}).
That is, the fictitious force that arises from moving into a non-inertial reference frame is canceled by the
background gravitational force.
Substituting the potential
$\Phi = \Phi_0 + \delta \Phi$
into the momentum equation, we thus find
\begin{multline}
    \rhoc \der{\vec{u}}{t}
=
    - \f{1}{a}\delx p_\mathrm{tot, c}
    - \rhoc \f{\dot{a}}{a} \vec{u}
    + \f{1}{a^2}\delx \bcdot \left(
    \vec{B}_\mathrm{c}\vec{B}_\mathrm{c} \right)
    - \rhoc \f{1}{a} \delx \delta \Phi \ ,
    \label{eq:mom-comoving-der}
\end{multline}
Finally, the remaining Poisson equation for $\delta \Phi$ is converted to comoving coordinates by defining the comoving total density,
$\rho_\mathrm{c, tot} = \rho_\mathrm{tot} a^3$. This gives
\be
    \delx^2 \delta \Phi
    = \f{4 \upi G}{a} \left(\rho_\mathrm{c, tot} - \bar{\rho}_\mathrm{c, tot}\right)\ .
\en
Sometimes the Poisson equation is further rewritten as (see e.g. \citealt{Weinberger2020})
\be
    \delx^2 \delta \Phi_\mathrm{c}
    = 4 \upi G \left(\rho_\mathrm{c, tot} - \bar{\rho}_\mathrm{c, tot}\right)\ .
\en
by defining the comoving potential as $\delta \Phi_\mathrm{c} = a \delta \Phi$ which is then related to
the full potential by
\be
    \delta \Phi_\mathrm{c} = a \Phi + \f{1}{2} a^2 \ddot{a} x^2 \ .
\en

\subsection{Induction equation}

Converting the time and space derivatives to the comoving coordinates, equation~\eqref{eq:ind} becomes
\be
    \pdertx{\vec{B}}
    =
    \f{\dot{a}}{a} \left(\vec{x}\bcdot \delx \right)\vec{B}
    +
    \f{1}{a}\delx \times \left(\vec{\varv}
    \times \vec{B}\right) \ .
\en
Substituting variables with comoving variables,
we obtain
\begin{multline}
    \pdertx{\vec{B}_\mathrm{c}}
    -
    \f{1}{a}\delx \times \left(\vec{u}
    \times \vec{B}_\mathrm{c}\right)
    = \\
    \f{\dot{a}}{a}\left[2\vec{B}_\mathrm{c} +
    \left(\vec{x}\bcdot \delx \right)\vec{B}_\mathrm{c} +
    \delx \times \left(\vec{x}
    \times \vec{B}_\mathrm{c}\right)
    \right] \ ,
    \label{eq:ind-comoving-derivation}
\end{multline}
Using a standard vector identity (see e.g. \citealt{Spruit2013}) we find
\be
    \delx \times \left(\vec{x}
    \times \vec{B}_\mathrm{c}\right)
    =
    -2 \vec{B}_\mathrm{c} - \left(\vec{x}\bcdot \delx \right)\vec{B}_\mathrm{c}
\en
which shows that the RHS of equation~\eqref{eq:ind-comoving-derivation} is zero. The comoving induction equation is therefore given by equation~\eqref{eq:Bc-comoving}.

\subsection{Internal energy equation}
We transform equation~\eqref{eq:internal-energy} by defining the comoving internal energy as $\epsc = \varepsilon a^{3}$. We substitute the comoving variables and obtain
\be
    \der{\ln \epsc}{t} - 3\der{\ln a}{t}
    =
    -\gamma \f{1}{a} \delx \bcdot \left(\dot{a} \vec{x} + \vec{u}\right) \ .
    \label{eq:epsc-inter}
\en
which, given that $\delx \bcdot \vec{x}=3$, reduces to equation~\eqref{eq:epcs-comoving}.

\subsection{Conservative form and total energy equation}

Cosmological MHD codes solve the comoving MHD equations in form which is as close to a set of conservation laws as possible. While equations~\eqref{eq:rho-comoving}--\eqref{eq:poisson-comoving}
are adequate for the analytic work presented here, we have found it
useful to connect our derivation to the existing literature.

We therefore briefly explain how to convert to
the form given in \citet{Pakmor2013}. The equations
presented here, however, include gravity and do not assume $\gamma=5/3$.
The derivation helped us understand how the extra energy source term that arises when $\gamma\neq5/3$ should be implemented in the \textsc{arepo} code.

The continuity equation, equation~\eqref{eq:rho-comoving}, can also be written as
\be
    \pdertx{\rhoc} + \f{1}{a} \delx \bcdot
    \left(\rho \vec{u} \right) = 0\ .
    \label{eq:rho-comoving-alternative}
\en
Equation~\eqref{eq:mom-comoving-der} can be transformed into an equation
for $\rhoc\vec{u}$ by writing out the convective derivative and
using the mass continuity equation. The momentum equation then becomes
\begin{multline}
    \pdertx{\rhoc \vec{u}}
    + \f{1}{a}\delx \bcdot
    \left[\rhoc \vec{u}\vec{u} +
    p_\mathrm{tot, c} \bb{1}
    -
    \f{\vec{B}_\mathrm{c}\vec{B}_\mathrm{c}}{a}
    \right]
    =
    - \rhoc \f{\dot{a}}{a} \vec{u}
    - \rhoc \f{1}{a} \delx \delta \Phi \ .
    \label{eq:mom-comoving-alternative}
\end{multline}
The total energy equation requires additional derivations.
As in \citet{Pakmor2013}, we define the total energy as
\be
    E_\mathrm{c} = \epsc + \f{1}{2}\rhoc u^2 + \f{\Bc^2}{2a} \ .
\en
The internal energy equation in conservative form becomes
\be
    \pdertx{\epsc} + \f{1}{a}\delx \bcdot \left(\epsc \vec{u} \right)
    =
    - \pc
    \f{1}{a}\delx \bcdot \vec{u}
    -3\f{\dot{a}}{a}\left(\gamma-1\right)\epsc \ .
    \label{eq:internal-comoving-conservative-form}
\en
The kinetic energy equation is found by dotting equation~\eqref{eq:mom-comoving-alternative}
with $\vec{u}$. After some manipulation one finds
\begin{multline}
    \left(\frac{\partial}{\partial t}\f{\rhoc u^2}{2} \right)_{\vec{x}}
    +
    \f{1}{a} \delx \bcdot \left[\f{\rhoc u^2}{2} \vec{u}
    + p_\mathrm{tot, c} \vec{u}
    - \f{\vec{B}_\mathrm{c} \left(\vec{u}\bcdot \vec{B}_\mathrm{c}\right)}{a}
    \right]
    = \\
    + \left(\pc + \f{\Bc^2}{2a}\right) \f{1}{a}\delx \bcdot \vec{u}
    - \f{1}{a^2}\vec{B}_\mathrm{c}\vec{B}_\mathrm{c}\bb{:} \delx\vec{u}
    - \rhoc \f{1}{a} \left(\vec{u} \bcdot \delx \right)\delta \Phi
    - \rhoc \f{\dot{a}}{a} u^2 \ ,
    \label{eq:kinetic-comoving}
\end{multline}
where the $\mathbf{:}$ notation is short-hand for the trace of
a matrix product (in index notation $\vec{A}\vec{B}\mathbf{:}\vec{C}\vec{D} = \sum_i\sum_j A_i B_j C_i D_j$).

The magnetic energy equation is found by dotting equation~\eqref{eq:Bc-comoving} with $\vec{B}_\mathrm{c}$. One finds
\be
    \left(\frac{\partial}{\partial t}\f{\Bc^2}{2} \right)_{\vec{x}}
    + \f{1}{a}\delx \bcdot \left(\f{\Bc^2}{2} \vec{u} \right)
    =
    - \f{1}{a} \left[\f{\Bc^2}{2} \delx \bcdot \vec{u}
    - \vec{B}_\mathrm{c}\vec{B}_\mathrm{c}\bb{:} \delx\vec{u}\right] \ ,
\en
which is straightforward to convert to an evolution equation for $\Bc^2/2a$. We find
\begin{multline}
    \left(\frac{\partial}{\partial t}\f{\Bc^2}{2a} \right)_{\vec{x}}
    + \f{1}{a}\delx \bcdot \left(\f{\Bc^2}{2a} \vec{u} \right)
    =
    - \f{\Bc^2}{2a} \f{1}{a} \delx \bcdot \vec{u}
    +\f{1}{a^2} \vec{B}_\mathrm{c}\vec{B}_\mathrm{c}\bb{:} \delx\vec{u}
    - \f{\Bc^2}{2a} \f{\dot{a}}{a} \ .
    \label{eq:magnetic-energy-comoving}
\end{multline}
The evolution equation for $E_\mathrm{c}$ is now found by adding up
equations~\eqref{eq:internal-comoving-conservative-form}, \eqref{eq:kinetic-comoving} and \eqref{eq:magnetic-energy-comoving}.
Most of the RHS terms cancel out as they correspond to transfer of energy
between thermal, kinetic and magnetic energy densities (as in standard MHD).
The equation for the comoving energy becomes
\begin{multline}
    \pdertx{E_\mathrm{c}} + \f{1}{a}\delx \bcdot \left[
    \left(E_\mathrm{c} +  p_\mathrm{tot, c} \right)\vec{u}
    - \f{\vec{B}_\mathrm{c} \left(\vec{u}\bcdot \vec{B}_\mathrm{c}\right)}{a}
    \right]
    = \\
    - \rhoc \f{1}{a} \left(\vec{u} \bcdot \delx \right)\delta \Phi
    -\left[3 \left(\gamma-1\right) \epsc + \rhoc u^2 + \f{\Bc^2}{2a}\right] \f{\dot{a}}{a} \ .
    \label{eq:comoving-energy}
\end{multline}

\subsection{Conservative form with fewer source terms}
The comoving MHD equations found in the literature
are equations for $a \rhoc \vec{u}$ and $a^2 E_\mathrm{c}$ rather
than simply $\rhoc \vec{u}$ and $E_\mathrm{c}$. The
reason is that this eliminates some of the cosmological source terms \citep{Pakmor2013}.

The conversion of the momentum equation is done by multiplying equation~\eqref{eq:mom-comoving-alternative}
by $a$ and rewriting the time derivative. This yields
\begin{multline}
    \pdertx{(a \rhoc \vec{u})}
    +
    \delx \bcdot \left[
    p_\mathrm{tot, c} \mathbf{1}
    +
    \rhoc \vec{u}\vec{u}
    - \f{1}{a}
    \vec{B}_\mathrm{c}\vec{B}_\mathrm{c}
    \right]
    =
    - \rhoc \delx \delta \Phi \ ,
\end{multline}
where the only remaining source term is the gravitational one.

It is similarly useful to consider the evolution of $a^2 E_\mathrm{c}$
rather than equation~\eqref{eq:comoving-energy}. The conversion is done by
multiplying by $a^2$ and rewriting the time derivative. This gives
\begin{multline}
    \pdertx{a^2E_\mathrm{c}} + a\delx \bcdot \left[
    \left(E_\mathrm{c} +  p_\mathrm{tot, c} \right)\vec{u}
    - \f{\vec{B}_\mathrm{c} \left(\vec{u}\bcdot \vec{B}_\mathrm{c}\right)}{a}
    \right]
    = \\
    - \rhoc a \left(\vec{u} \bcdot \delx \right)\delta \Phi + \left[\left(5 - 3 \gamma\right) \epsc + \f{\Bc^2}{2a}\right] a\dot{a}
    \label{eq:comoving-energy-with-fewer-sources} \ .
\end{multline}
For $\gamma=5/3$ there is only a single (magnetic) cosmological source term in equation~\eqref{eq:comoving-energy-with-fewer-sources}. When $\gamma\neq5/3$ there is an additional source term related to the thermal energy. We have implemented this term in \textsc{arepo} in order to perform
simulations with $\gamma=4/3$ (i.e. Figures~\ref{fig:compressible-gamma-4/3-standing} and \ref{fig:compressible-gamma-4/3-traveling}).\footnote{The source term vanishes for
$\gamma=5/3$ when the comoving internal energy is defined as $\epsc = \varepsilon a^3$. We have found it practical to keep this conventional definition of $\epsc$. We note, however, that it is always possible to eliminate the source term in equation~\eqref{eq:comoving-energy-with-fewer-sources} by instead defining $\epsc$ as
$\epsc = \varepsilon a^{3\gamma-2}$.}.

\section{Analytic solution details}
\label{app:analytic-details}

\subsection{Alfvén solution for $\Oma=1/4$}
\label{app:Oma=1/4}

Equation~\eqref{eq:alfven-Bc-gen-sol} is not the solution to
equation~\eqref{eq:alfven-ODE} in the special case $\Oma=1/4$
(which gives $\kappa=0$). Instead the solution is
\be
     \f{\dBc}{\Bc} = a^{-1/4} \left(c_3 + c_4 \ln a\right) \ ,
\en
where $c_3$ and $c_4$ are integration constants \citep{asmar2010partial}. The change in solution occurs because the indicial roots of equation~\eqref{eq:alfven-ODE} are not distinct. This particular value of $\Oma$ does not correspond to a wave but we give its solution for completeness. Applying the initial conditions, we find
\be
    \f{\dBc}{\Bc}
    =
    \left(\f{a}{\ai}\right)^{-1/4}
    \left[
    A_B
    + \f{A_B + \ui\sqrt{\ai}A_u}{4}
    \ln\left(\f{a}{\ai}\right)
    \right]\ ,
    \\
    \f{\du}{\vao} =
    \left(\f{a}{\ai}\right)^{-3/4}
    \left[
    A_u
    -
    \f{A_u - \ui \ai^{-1/2} A_B}{4}
    \ln\left(\f{a}{\ai}\right)
    \right] \ .
\en

\subsection{Compressible solution for $\gamma=4/3$ with $\sigma=1/4$}
\label{app:sigma=1/4}

The case $\sigma=1/4$ in equation~\eqref{eq:ODE-gamma-4/3} requires special treatment for the same reason as
in Appendix~\ref{app:Oma=1/4} above. Solving and applying initial conditions, we find that the compressible solution in this case becomes

\be
    \f{\drhoc}{\rhoc} =
    \left(\f{a}{\ai}\right)^{-1/4}
    \left[
    A_\rho
    + \f{A_\rho - 4\ui\Oms \sqrt{\ai}A_u}{4}
    \ln\left(\f{a}{\ai}\right)
    \right]\ , \\
    \f{\du}{\cado} =
    \left(\f{a}{\ai}\right)^{-3/4}
    \left[
    A_u
    -
    \left(\f{A_u}{4}
    +  \f{\ui A_\rho}{16\sqrt{\ai}\Oms}
    \right)
    \ln\left(\f{a}{\ai}\right)
    \right] \ .
\en

\subsection{Comoving magnetosonic wave with self-gravity}
\label{app:magnetosonic-derivation-details}

This appendix provides additional details on the derivation
presented in Section~\ref{sec:comp-gamma-not-4/3}.
We compare our equation~\eqref{eq:rhoc-ODE-comp}
with equation 6.80 in §104 in \citet{Bowman1958}. The latter equation
reads\footnote{The overlap in notation can be somewhat confusing
and we have for this reason put a prime on $\gamma'$ in
equation~\eqref{eq:bowman-ode}.}
\be
    \dder{y}{x} - \f{2\alpha-1}{x} \der{y}{x}
    +
    \left(
    \beta^2 \gamma'^2 x^{2\gamma'-2}
    + \f{\alpha^2 - n^2 \gamma'^2}{x^2}
    \right)y = 0 \ .
    \label{eq:bowman-ode}
\en
Setting $y=\drhoc/\rhoc$ and
$x=a$ the comparison immediately reveals that $\alpha=-1/4$
is required in order to match the $dy/dx$ term.
Next, we realize that we need
$
    2 \gamma' - 2 = 2 - 3\gamma
$
 in order to have the same power of $x$ in the
term proportional to $\beta^2$. This leads us to
define $s = \gamma' \equiv (4-3\gamma)/2$ as in equation~\eqref{eq:s-parameter}.
We also require
$
    \beta^2 s^2 = \Oms^2
$
so that $\beta = \Oms/|s|$.
Finally we observe that
$
    \alpha^2 - n^2 \gamma'^2 = \Oma^2 - \Omg^2
$
from which we find
\be
    s^2 n^2 = \f{1}{16} - \left(\Oma^2 - \Omg^2\right) \ ,
\en
which leads to our definition of $\nu=n$ given in equation~\eqref{eq:nu-parameter}.
According to equation 6.82 in \citet{Bowman1958} the solution to
equation~\ref{eq:bowman-ode} valid also for integer $n$ is
\be
    y = x^{\alpha} \left\{A\, J_n\left(\beta x^{\gamma'}\right)
                    + B\, Y_n\left(\beta x^{\gamma'}\right)\right\}\ ,
\en
where $A$ and $B$ are integration constants. The solution to equation~\eqref{eq:rhoc-ODE-comp} is thus
given by equation~\eqref{eq:drho-besselY-solution}
where $\mathcal{F}(a)$ and $\mathcal{G}(a)$ are defined in
equation~\eqref{eq:FandG}.

The integration constants, $c_1$ and $c_2$, are found by solving two
equations for two unknowns: $\drhoc(\ai)/\rhoc = c_1 \mathcal{F}(\ai) + c_2 \mathcal{G}(\ai)$ and $\drhoc'(\ai)/\rhoc = c_1 \mathcal{F}'(\ai) + c_2 \mathcal{G}'(\ai)$.
This yields
\be
    c_1 = \f{\mathcal{G}'(\ai)}{D(\ai)}\f{\drhoc}{\rhoc}(\ai) - \f{\mathcal{G}(\ai)}{D(\ai)}\f{\drhoc'}{\rhoc}(\ai) \ ,
\en
\be
    c_2 = \f{\mathcal{F}(\ai)}{D(\ai)}\f{\drhoc'}{\rhoc}(\ai) - \f{\mathcal{F}'(\ai)}{D(\ai)}\f{\drhoc}{\rhoc}(\ai) \ ,
\en
where
\be
    D(\ai) = \mathcal{F}(\ai) \mathcal{G}'(\ai) - \mathcal{F}'(\ai)\mathcal{G}(\ai) = \f{2 s}{\upi \ai^{3/2}} \ .
\en
We simplify these expressions by
defining dimensionless amplitudes $A_u = \du(\ai)/\cado$ and $A_\rho = \drhoc(\ai)/\rhoc$
and using equation~\eqref{eq:drhoc-da-comp-final} to write
\be
    \f{\drhoc'}{\rhoc}(\ai) = - \f{\ui k}{H_0 \sqrt{\ai}}\du(\ai)
    = - \f{\ui \Oms}{\sqrt{\ai}}A_u \ .
\en
In terms of these,
the integration constants are given by
\be
    c_1
    &=&
    \;\,\,\,\f{\upi \ai^{3/2}}{2 s}\left(\mathcal{G}'(\ai)A_\rho  + \ui \Oms
    \f{\mathcal{G}(\ai)}{\sqrt{\ai}} A_u\right)
    \label{eq:c1-magnetosonic-general-gamma}
    \ , \\
    c_2
    &=&
    -
    \f{\upi \ai^{3/2}}{2 s}\left(
        \mathcal{F}'(\ai)A_\rho
        +
        \ui \Oms \f{\mathcal{F}(\ai)}{\sqrt{\ai}}A_u
    \right)
    \label{eq:c2-magnetosonic-general-gamma}
    \ .
\en
For practical computations we use the following explicit expressions for the derivatives of $\mathcal{F}(a)$ and $\mathcal{G}(a)$:
\be
    \mathcal{F}'(a) = \Oms\sgn(s) a^{s-5/4}  J'_\nu\left(z\right)
                   -\f{1}{4}a^{-5/4}     J_\nu\left(z\right) \ , \\
    \mathcal{G}'(a) = \Oms\sgn(s) a^{s-5/4}  Y'_\nu\left(z\right)
                   -\f{1}{4}a^{-5/4}     Y_\nu\left(z\right) \ ,
\en
where $\sgn(s)$ is the sign function.

\subsection{Alfvén wave including Navier-Stokes viscosity}
\label{app:alfven-with-visc}

We provide a derivation of the analytic solution for an Alfvén wave
including physical Navier-Stokes viscosity. This solution is used to
interpret the numerical dissipation study presented in
Fig.~\ref{fig:alfven-sim-standing}. The analysis is a generalization
of the first part of Section~\ref{sec:analytic-alfven-wave} in the main text.
Here we limit ourselves to standing waves and hence do not provide the
expressions for traveling waves with viscosity (mainly due to space
restrictions).
We also note that it is
possible to derive analytic results including viscosity for
compressible waves with
$\gamma=4/3$. We do not pursue this here but instead refer to Berlok et al, in prep, for such a calculation including Braginskii viscosity.

Navier-Stokes viscosity is included as an
additional term on the RHS of equation~\eqref{eq:mom},
$-\delr \bcdot \mathbf{\Pi}$, where
\be
    \mathbf{\Pi} = \eta \left(\delr \vec{\varv} + \left(\delr \vec{\varv}\right)^T
    - \f{2}{3} \mathbf{1} \delr \bcdot \vec{\varv} \right) \ ,
\en
is the viscosity tensor and $\eta$ is the viscosity coefficient.
After transforming to comoving coordinates, this leads to an
additional term,
\be
    - a^2 \eta \delx \bcdot \left(\delx \vec{u} + \left(\delx \vec{u}\right)^T
    - \f{2}{3} \mathbf{1} \delx \bcdot \vec{u} \right) \ ,
\en
on the RHS of equation~\eqref{eq:mom-comoving}. We will assume that the
viscosity coefficient is given by $\eta = \eta_0 a^n$ where $\eta_0$
is its value at $z=0$. We further set $n=-5/2$ since this leads to an
analytically solvable ODE. We also define a viscosity parameter
\be
    \Gamma = \f{\eta_0 k^2}{\rhoc H_0} \ ,
\en
which simplifies the following results.
The new linearized equations for an Alfvén wave,
corresponding to equations~\eqref{eq:Bc-alfven} and \eqref{eq:a-du-alfven},
then become
\be
    \pder{}{a}\f{\dBc}{\Bc} = \f{\ui\Oma}{a^{3/2}} \f{\dw}{\vao}
    \ ,
    \label{eq:Bc-alfven-visc} \\
    \pder{}{a}\left(\f{\dw}{\vao}\right) = \f{\ui \Oma}{a^{1/2}} \f{\dBc}{\Bc}
    - \f{\Gamma}{a} \f{\dw}{\vao}
    \label{eq:a-du-alfven-visc}
    \ ,
\en
where $\dw = a \du$. We combine these linearized equations and obtain
\be
\pdder{}{a}\left(\f{\dw}{\vao}\right) + \f{1+2\Gamma}{2a} \pder{}{a}\left(\f{\dw}{\vao}\right) + \f{\Oma^2 - \Gamma/2}{a^{2}} \f{\dw}{\vao} = 0 \ ,
\en
which is solved using the usual machinery \citep{asmar2010partial}. We find
the solution for $\du=\dw/a$ to be
\be
\f{\du}{\vao} = a^{-3/4-\Gamma/2} \left(c_1 e^{i \kappa \ln a}
+ c_2 e^{-i \kappa \ln a}\right) \ ,
\en
where
\be
    \kappa \equiv \sqrt{\Oma^2 - \f{\left(1+2\Gamma\right)^2}{16}} \ ,
    \label{eq:alfven-kappa-visc}
\en
and $c_1$ and $c_2$ are integration constants. The solution for $\dBc$ is
found by integrating equation~\eqref{eq:Bc-alfven-visc} and is given by
\be
\f{\dBc}{\Bc} = -\ui
a^{-1/4-\Gamma/2}
\left(
c_3e^{i\kappa \ln a} +
c_4e^{-i\kappa \ln a}
\right) \ ,
\en
where $c_3 = 4\Oma c_1/(1 + 2 \Gamma - 4\ui\kappa)$ and
$c_4 = 4\Oma c_2/(1 + 2 \Gamma + 4\ui\kappa)$.
The solution with $\dBc(\ai)/\Bc = A_B$ and $\du(\ai)/\vao=A_u$
is given by
\begin{multline}
    \f{\dBc}{\Bc}
    =
    \left(\f{a}{\ai}\right)^{-1/4-\Gamma/2}
    \Big[
    A_B \cos(\psi)
    + \Big.\\ \Big.
    \f{A_B\left(1 + 2 \Gamma \right) + 4\ui\Oma \sqrt{\ai}A_u}{4\kappa}
    \sin(\psi)
    \Big]\ ,
    \label{eq:analytic-alf-dBc-osc-visc}
\end{multline}
\begin{multline}
    \f{\du}{\vao} =
    \left(\f{a}{\ai}\right)^{-3/4-\Gamma/2}
    \Big[
    A_u \cos\left(\psi\right)
    - \Big.\\ \Big.
    \f{A_u\left(1 + 2 \Gamma \right) - 4 \ui \Oma \ai^{-1/2} A_B}{4\kappa}
    \sin\left(\psi\right)
    \Big] \ ,
    \label{eq:analytic-alf-du-osc-visc}
\end{multline}
where $\psi = \kappa \ln (a/\ai)$ and $\kappa$ is given by
equation~\eqref{eq:alfven-kappa-visc} which includes the viscosity effect.
These equations reduce to equations~\eqref{eq:analytic-alf-dBc-osc} and \eqref{eq:analytic-alf-du-osc} in the limit of zero viscosity where $\Gamma=0$.
Viscosity modifies the wave by increasing the damping
(by a factor of $a^{-\Gamma/2}$) and decreasing the frequency (see equation
\ref{eq:alfven-kappa-visc}). It also changes the criterion for when oscillating
waves are possible, i.e., this criterion becomes $\Oma > \left(1 + 2 \Gamma \right)/4$
rather than simply $\Oma > 1/4$. We note that the transition from
oscillating waves to pure damping, i.e., $\kappa = 0$, requires special
treatment (as in Appendix~\ref{app:Oma=1/4}) and has a solution that differs
from
Equations~\eqref{eq:analytic-alf-dBc-osc-visc} and \eqref{eq:analytic-alf-du-osc-visc}. We do not pursue this here.


\bsp
\label{lastpage}
\end{document}